\documentclass[twocolumn,10pt,showkeys,preprintnumbers,amsmath,amssymb]{revtex4}
\usepackage{graphicx}
\usepackage{dcolumn}
\usepackage{bm}
\usepackage{amsmath}
\usepackage{enumerate}
\newcommand {\me} {e}

\newcommand {\dif} {d}
\newcommand {\mj} {j}
\begin{document}

\title{Statistics of Impedance and Scattering Matrices \\in  Chaotic
Microwave Cavities: Single Channel Case}  
\author{Xing Zheng}
\author{Thomas M. Antonsen Jr. }
\altaffiliation[Also at ] {Department of Electrical and Computer
Engineering.}
\author{Edward Ott}
\altaffiliation[Also at ] {Department of Electrical and Computer
Engineering.}\affiliation{Department of Physics\\ and Institute
for Research in Electronics and Applied Physics,\\ University of
Maryland, College Park, MD, 20742}
\date{\today}

\begin{abstract}
We study the statistical properties of the
impedance ($Z$) and scattering ($S$) matrices of open
electromagnetic cavities with several transmission lines or
waveguides connected to the cavity. In this paper, we mainly discuss
the single port  case. The generalization to multiple ports  is treated in
a companion paper. The model we consider is based on assumed
properties of chaotic eigenfunctions for the closed system.
Analysis of the model successfully reproduces features of the
random matrix model believed to be universal, while at the same
time incorporating features which are specific to individual
systems as treated by the Poisson kernel of Mello et al. Statistical
properties of the cavity impedance $Z$ are
obtained in terms of the radiation impedance (i.e., the impedance
seen at a port with the cavity walls moved to infinity). Effects
of wall absorption are discussed. Theoretical predictions are tested by
direct comparison with numerical solutions for a specific system.
(Here the word universal is used to denote high frequency statistical
properties that are shared by the members of the general class of systems
whose corresponding ray trajectories are chaotic. These universal
properties are, by definition, independent of system-specific details.)

 \end{abstract}


\maketitle                 

\newpage

\section{\label{sec:level1}Introduction} 

The problem of the coupling of electromagnetic radiation in and
out of structures is a general one which finds applications in a
variety of scientific and engineering contexts.  Examples include
the susceptibility of circuits to electromagnetic interference,
the confinement of radiation to enclosures, as well as the
coupling of radiation to  structures used to accelerate charged particles.

Because of the wave nature of radiation, the coupling properties
of a structure depend in detail on the size and shape of the
structure, as well as the frequency of the radiation.  In
considerations of  irregularly shaped electromagnetic enclosures
for which the wavelength is fairly small compared with the size of
the enclosure, it is typical that the electromagnetic field
pattern within the enclosure, as well as the response to external
inputs, can be very sensitive to small changes in frequency and to
small changes in the configuration. Thus, knowledge of the
response of one configuration of the enclosure may not be useful
in predicting that of a nearly identical enclosure.  This
motivates a statistical approach to the electromagnetic problem.

While our ability to numerically compute the response of
particular structures has advanced greatly in recent years, the
kind of information needed for a statistical description may not
be obtainable directly from numerical computation.

Thus it would seem to be desirable to have specific analytical
predictions for the statistics of electromagnetic quantities in
such circumstances.  This general problem has received much
attention in previous work (e.g.,
Refs.~\cite{holland99,lehman91,kostas91,price93,holland94,hill98,
cappetta98}). Some of the main issues addressed are:\ \ the
probability distribution of fields at a point, the correlation
function of fields at two points near each other, the statistics
of the excitation of currents in cables or in small devices within
the enclosure, the cavity $Q$, the statistics of coupling to the
enclosure, and the statistics of scattering properties.  A
fundamental basis for most of these studies is that, due to the
complexity of the enclosure and the smallness of the wavelength
compared to the enclosure size, the electromagnetic fields
approximately obey a statistical condition that we shall call {\it
the random plane wave hypothesis}, which assumes that a
superposition of random plane wave can be used to describe the
statistics of chaotic wave functions \cite{berry83}.
This work has been quite successful in obtaining meaningful
predictions, and some of these have been tested against
experiments with favorable results.  A good introduction and
overview is provided in the book by Holland and St. John
\cite{holland99}.

In addition to this previous work on {\it statistical
electromagnetics}
\cite{holland99,lehman91,kostas91,price93,holland94,hill98,
cappetta98}, much related work has been done by theoretical
physicists. The physicists are interested in solutions of quantum
mechanical wave equations when the quantum mechanical wavelength
is short compared with the size of the object considered.  Even
though the concern is not electromagnetics, the questions
addressed and the results are directly applicable to wave
equations, in general, and to electromagnetics, in particular. The
start of this line of inquiry was a paper by Eugene Wigner
\cite{wigner58}.  Wigner's interest was in the energy levels of
large nuclei.  Since the energy level density at high energy is
rather dense, and since the solution of the wave equations for the
levels was inaccessible, Wigner proposed to ask statistical
questions about the levels. Wigner's results apply directly to the
statistics of resonant frequencies in highly-overmoded
irregularly-shaped electromagnetic cavities. Since Wigner's work,
and especially in recent years, the statistical approach to wave
equations has been a very active area in theoretical physics,
where the field has been called `quantum chaos'. We emphasize,
however, that the quantum aspect to this work is not inherent, and
that a better terminology, emphasizing the generality of the
issues addressed, might be `wave chaos'. For a review see Chapter
11 of Ref. \cite{ott02} or the books \cite{gutz90,haake91}.

Wigner's approach was to introduce what is now called Random Matrix
Theory
(RMT) \cite{mehta91}. In RMT the linear wave equation is replaced or
modelled by a linear matrix equation where the elements of the matrix are
random variables. This follows from Wigner's hypothesis that the
eigenvalues for a complicated (in our case chaotic) system have the same
statistics as those of matrices drawn from a suitable ensemble. Based on
symmetry arguments, Wigner proposed that the matrix statistics are those
that would result if the matrix were drawn from different types of
ensembles, where the relevant ensemble type depends only on gross
symmetries of the modelled system. The two ensembles that are relevant to
electromagnetic problems are the Gaussian Orthogonal Ensemble (GOE) and
the Gaussian Unitary Ensemble (GUE). In both cases, all the matrix
elements are zero mean Gaussian random variables. In the GOE all the
diagonal element distributions have the same width, while all
the off diagonal element distributions have widths that are half that of
the diagonal elements. The matrices are constrained to be symmetric, but
otherwise the elements are statistically independent. The GOE case is
intended to model wave systems that have time reversal symmetry
(TRS). That is, the time domain equations are invariant under the
transformation $t\rightarrow -t$. This is the case for electromagnetic
waves if the permitivities and permeabilities tensors are real and
symmetric. In
the GUE the matrices are constrained to be {\it Hermitian}. In this case
the off-diagonal  elements are complex and the distributions of their
real and imaginary parts are independent and Gaussian and the width of
these Gaussians is again the one half the width of the real diagonal
elements. The GUE case is intended  to model systems for which time
reversal symmetry is broken (TRSB). This case will apply in
electromagnetics if a nonreciprocal  element such as a magnetized ferrite
 or a cold magnetized plasma, is added to
the system.

In this paper we mainly consider an irregularly shaped cavity with
a single transmission line and/or waveguide connected to it, and we
attempt to obtain the statistical properties of the impedance  $Z$ and
the scattering matrix $S$ (which are both
scalars in the cases we consider) characterizing the
response of the cavity to excitations from the connected
transmission line,  where the
wavelength is small compared to the size of the cavity.  We will
treat specifically the case of cavities that are thin in the vertical
($z$-direction) direction.  In this case the resonant fields of
the closed cavity are transverse electromagnetic ($TM_z$, ${\vec
E}=E_z(x,y){\hat z})$, and the problem admits a purely scalar
formulation.  While the two dimensional problem has practical
interest in appropriate situations (e.g., the high frequency
behavior of the power plane of a printed circuit), we emphasize
that the results for the statistical properties of $Z$ and $S$
matrices are predicted to apply equally well to three dimensional
electromagnetics and polarized waves. We note that previous work
on statistical electromagnetics
\cite{holland99,lehman91,kostas91,price93,holland94,hill98,
cappetta98} is for fully three dimensional situations.  Our main
motivation for restricting our considerations here to two
dimensions is that it makes possible direct numerical tests of our
predictions (such numerical predictions might be problematic in
three dimensions due to limitations on computer capabilities).
Another benefit is that analytical work and notation are
simplified.

 For an electrical circuit or electromagnetic cavity with
ports, the impedance matrix provides a characterization of the structure in
terms of the linear relation between the voltages and currents at
all ports,
\begin{equation}
\hat{V}=Z\hat{I},
\label{eq:defineZ}
\end{equation}
where $\hat{V}$ and $\hat{I}$ are column vectors of the complex phasor
amplitudes of the sinusoidal port voltages and currents.
The scattering matrix $S$ is related to the impedance matrix $Z$ by
\begin{equation}
S=Z_0^{1/2}(Z+Z_0)^{-1}(Z-Z_0)Z_0^{-1/2},
\end{equation}
where $Z_0$ is the characteristic impedance of the transmission line.

As discussed in the next section, the impedance matrix $Z$ can be
expressed in terms of the eigenfunctions and eigenvalues of the
closed cavity. We will argue that the elements of the $Z$ matrix
can be represented as combinations of random variables with
statistics based on the random plane wave hypothesis for the
representation of chaotic wave functions and results from random
matrix theory
\cite{wigner58, ott02} for the distribution of the
eigenvalues.

This approach to determination of the statistical properties of
the $Z$ and $S$ matrices allows one to include the  generic
properties of these matrices, as would be predicted by
representing the system as a random matrix drawn from an approapriate
ensemble. It also, however, allows one to treat aspects of the $S$
and $Z$ matrices which are specific to the problems under
consideration (i.e., so-called \emph{non-universal} properties).

These nonuniversal properties have previously been treated within the
context of the so-called Poisson
Kernel based on a ``maximum information entropy" principle
\cite{mello85},
and Brouwer later provided a microscopic justification and showed that
the Poisson Kernel can be derived from Wigner's RMT description of
the Hamilton \cite{brouwer95}. Here the statistics of the $S$ matrix
depend in a non-trivial way on the average of $S$ taken  in a narrow
frequency range. This characterizes the system specific aspects of the
coupling. Our approach allows one to
{\it predict} the average based on another
informative quantity, the  radiation impedance, which itself
characterizes the coupling of the port to the enclosure.
The radiation impedance is the impedance that applies at the port when waves
are launched into the cavity and (by making the distant walls perfectly
absorbing) not allowed to return. Our interpretation of the role of the
radiation impedance is equivalent to Brouwer's \cite{brouwer95}
interpretation of the Poisson kernel in terms of scattering from a cascaded
configuration of a lossless multiport and a perfectly coupled cavity
described by Random Matrix Theory. We show that the separation of universal
and system specific properties is more natural when considering the
impedance rather than scattering matrix. Specifically, the universal
properties of the cavity impedance are observed by subtracting from
the raw cavity impedance the radiation reactance and normalizing the result
to the radiation resistance. The statistics of the resulting variable, which
we term the normalized cavity impedance, depend only on a single parameter
characterizing the internal loss of the cavity.

The Poisson kernel description of the scattering amplitude has
been applied to data obtained from microwave scattering
experiments on cavities with absorption \cite{smilansky, 
stockmann1, stockmann2}. Two different methods have recently been described
\cite{stockmann1, stockmann2} for extracting the universal
properties of the scattering amplitude from the system specific
ones. The system specific properties are characterized in terms of
averages of the scattering amplitude and its modulus squared over
ranges of frequency. A fitting procedure is then used to
characterize the universal fluctuations. The impedance approach
has also been applied to experimental data \cite{anlageprl,
warne03}. In ref.~\cite{anlageprl} the radiation impedance is
measured directly and the complex normalized  impedance is formed
by subtracting the radiation reactance and normalized to the
radiation resistance. The probability distribution function of the
normalized impedance was then compared directly with Monte Carlo
numerical evaluations of the prediction.

The main contribution of this paper are to describe the
statistical properties of the cavity impedance. First we show
(also shown in Ref.~\cite{warne03}) that the relation between the
cavity impedance and the radiation impedance follows from the
assumption that the eigenfunctions of the cavity satisfy the
random plane wave hypothesis of Berry \cite{berry83}. We then
verify the relation between the cavity impedance and the radiation
impedance for a specific realization by numerical simulation using
the High Frequency Structure Simulation software, HFSS. Next,
using Monte Carlo methods we evaluate theoretical predictions for
the probability distribution functions for the real and imaginary
parts of the normalized impedance for different values of internal
absorption in the cavity. The distribution of the real part of the
normalized impedance is closely related to the distribution of
values that is known as the local density of states \cite{efetov}.
Finally, we derive expressions for the mean and variance of the
normalized impedance as functions of the level of internal loss in
the cavity.

 Our paper is organized as
follows.  In Sec.~II, we presents the statistical model. Section
III illustrates our model by application to the statistics of the
impedance seen at a single transmission line input to a cavity
that is irregularly-shaped, highly over-moded, lossless, and
non-gyrotropic (i.e., no magnetized ferrite).  Section IV relates
the impedance matrix characteristics to those of the scattering
matrix. Section V generalizes our model to incorporate the effects
of distributed loss (such as wall absorption). Throughout, our
analytical results will be compared with direct numerical
solutions of the wave problem. Section VI concludes with a
discussion and summary of results.

\section{Modelling with Random Plane Waves}

\begin{figure}
\includegraphics[scale=0.7]{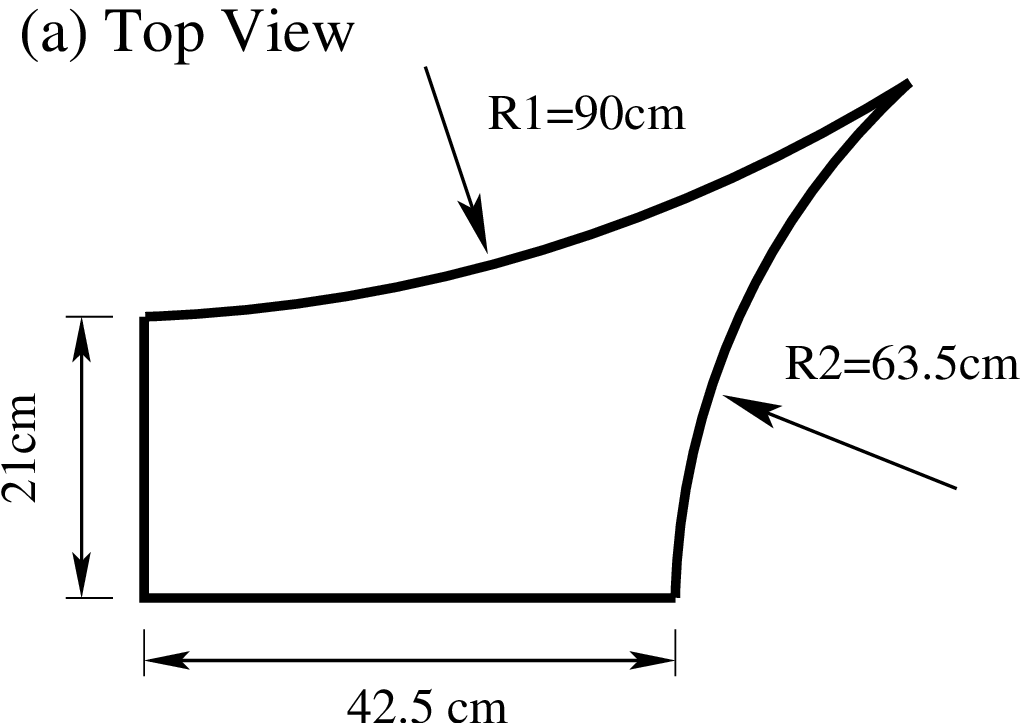}
\includegraphics[scale=0.7]{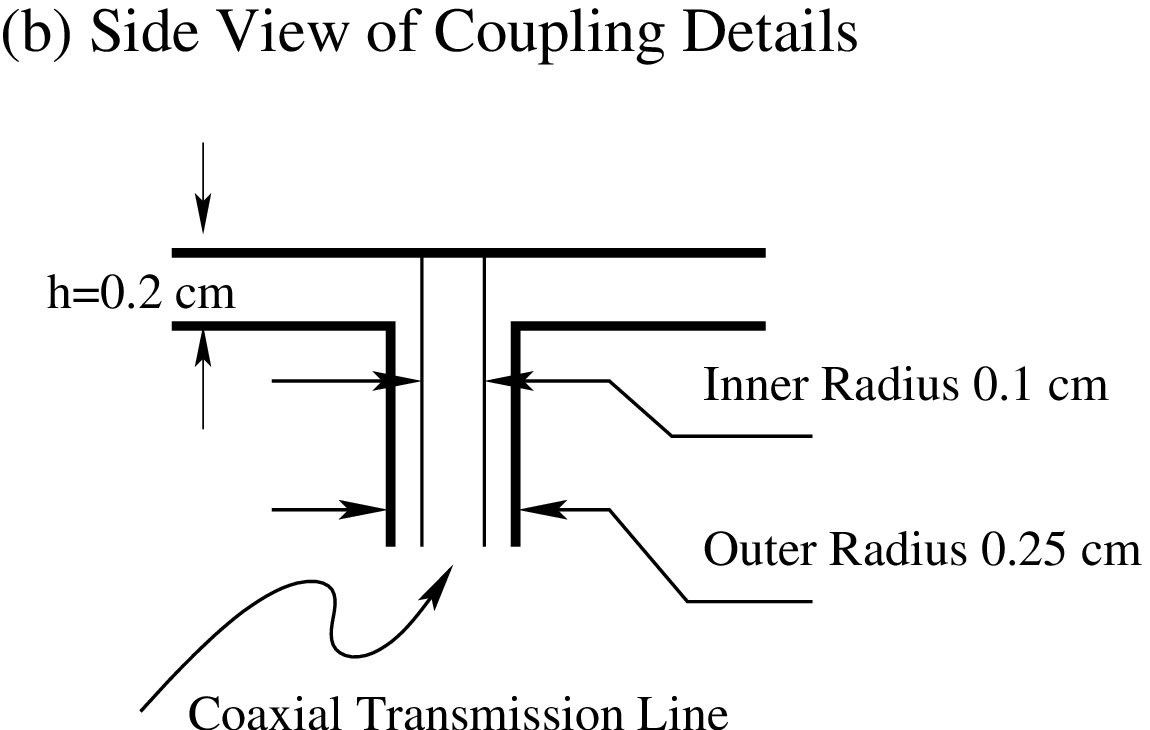}
\caption{(a) Top view of the cavity used in our numerical
simulation. (b) Side view of the details of a possible coupling.}
\label{fig:bowtie}
\end{figure}

We consider a closed  cavity   with ports connected to it. For
specificity, in our numerical work, we consider the particular,
but representative, example of the vertically thin cavity shown in
Fig.~\ref{fig:bowtie}(a) coupled to the outside via a coaxial
transmission cable. Fig.~\ref{fig:bowtie}(b) shows an example of
how this cavity might be connected to a transmission line via a
hole in the bottom plate. The cavity shape in
Fig.~\ref{fig:bowtie} is of interest here because the concave
curvature of the walls insures that typical ray trajectories in
the cavity are chaotic. (Fig.~\ref{fig:bowtie}(a) is a quarter of
the billiard shown in Fig.~2(c).) For our purposes, a key consequence of
the chaotic property of the shape in Fig~1(a) is that, if we consider
the trajectory of a particle bouncing with specular reflection at
the walls (equivalently a ray path), then a randomly  chosen
initial condition (i. e., random in position $\vec{x}$ within the
cavity and isotropically random in the orientation $\theta$ of the
initial velocity vector) always generates an orbit that is {\it
ergodic} within the cavity. Here by ergodic we mean the following:
For any spatial region $R$ within the cavity, in the limit of time
$t\rightarrow \infty$, the fraction of time the orbit spends in
$R$ is the ratio of the area of $R$ to the entire area of the
two-dimension cavity, and, furthermore, the collection of velocity
orientations $\theta$ of the orbit when it is in $R$ generates a
uniform distribution in [0, 2$\pi$). Thus the orbit {\it
uniformly} covers the phase space ($\vec{x}$, $\theta$). This is
to be contrasted with the case of a rectangular cavity, $0\le x
\le a$, $0 \le y \le b$, which represents a shape for which orbits
(rays) are nonchaotic. In that case, if the initial velocity
orientation with respect to the x-axis is $\theta_0$ then at any
subsequent time only four values of $\theta$ are possible:
$\theta_0$, $2\pi-\theta_0$, $\pi-\theta_0$, $\pi+\theta_0$. In
cases such as Fig.~1(a) we assume that the previously mentioned
hypotheses regarding eigenfunctions and eigenvalue distributions
provide a useful basis for deducing the statistical properties of
the $Z$ and $S$ matrices, and, in what follows, we investigate and
test the consequences of this assumption.

The vertical height $h$ of the cavity is small, so that, for
frequencies of interest, the only propagating waves   inside the
cavity have electric fields that are purely vertical,
\begin{equation}
\vec{E}=E_{z}(x,y)\hat{z}.
\end{equation}
This electric field gives rise to a charge density on the top plate
$\rho_s=-\epsilon_0 E_{z}$, and also generates a voltage
$V_T(x,y)=-hE_z(x,y)$ between the plates. The magnetic
field is perpendicular to $\hat{z}$,
\begin{equation}
\vec{B}=(B_x,B_y)=\mu_0 \vec H,
\end{equation}
and is associated with a  surface current density
$\vec{J}_s=\vec{H}\times\hat{z}$ flowing on the top plate.

The cavity excitation problem for a geometry like that in
Fig.~\ref{fig:bowtie}(b) is system specific. We will be interested
in separating out statistical properties that are independent of
the coupling geometry and have a universal (i.e.,
system-independent) character. For this purpose, we claim that it
suffices to consider a simple solvable excitation problem, and
then generalize to more complicated cases, such as the coupling
geometry in Fig.~\ref{fig:bowtie}(b). Thus we consider the closed
cavity (i.e., with no losses or added metal), with localized
current sources  resulting in a current density
$\vec{J}_s(x,y,t)=\sum_i I_i(t) u_i(x,y)\hat{z}$ between the
plates. The profile functions $u_i(x,y)$ are assumed to be
localized; i.e., $u_i(x,y)$ is essentially zero for
$(x-x_i)^2+(y-y_i)^2>l_i^2$, where $l_i$ is much smaller than the
lateral cavity dimension. $u_i(x,y)$ characterizes the
distribution of vertical current at the location of the $i$-th
model input (analogous to the $i$-th transmission line connected
to the cavity, although, for this model there are no holes in the
upper or lower plates). The profile is normalized such that
\begin{equation}
\int \dif x \dif y u_i(x,y)=1.
\label{eq:profileu}
\end{equation}

For the sake of simplicity, we only consider the single port case
in this paper (i.e., there is only one localized source and we may thus
drop the subscript $i$ on $u_i(x,y))$.
The injection of current serves as a source in
the continuity equation for surface charge, $\partial
\rho_s/\partial t+\nabla \cdot\vec{J_s}=Iu(x,y)$, where $\nabla
=(\partial/\partial x,\partial/\partial y)$. Expressed in terms of
fields, the continuity equation becomes:
\begin{equation}
\frac{\partial}{\partial t}(-\epsilon_{0}E_{z})+\nabla
\cdot(\tilde{H}\times\hat{z})=Iu(x,y). \label{eq:continuity}
\end{equation}
Differentiating Eq.~(\ref{eq:continuity}) with respect to $t$ and using
Faraday's law, we obtain,
\begin{equation}
\frac{\partial^2}{\partial t^2}(-\epsilon_{0}E_{z})+\nabla \cdot
\frac{1}{\mu_0}\nabla E_z=u(x,y) \frac{\partial I}{\partial t}.
\end{equation}
Expressing the electric field in terms of the voltage $V_T=-E_zh$,
we arrive at the driven wave equation,
\begin{equation}
\frac{1}{c^2}\frac{\partial^2}{\partial t^2}V_T-\nabla ^2
V_T=h\mu_0 u \frac{\partial I}{\partial t},
\end{equation}
where $c$ is speed of light, $c^2=1/(\mu_0 \epsilon_0)$.

Assuming sinusoidal time dependence $e^{j\omega t}$ for all field
quantities, we obtain the following equation relating $\hat{V}_T$
and $\hat{I}$,  the phasor amplitudes of the voltage between the
plates and the  port current,
\begin{equation}
(\nabla ^2+k^2)\hat{V}_T=-\mj \omega h \mu_{0}u\hat{I}=-\mj kh
\eta_0 u\hat{I}, \label{eq:virelation}
\end{equation}
where $\eta_0=\sqrt{\mu_0/\epsilon_0}$ is the characteristic
impedance of free space and $k=\omega/c$. Thus
Eq.~(\ref{eq:virelation}) represents
a wave equation for the voltage between the plates
excited by the input current.

To complete our description and arrive at an expression of the form of
Eq.~(\ref{eq:defineZ}), we need to determine the port voltage $V$. We
take its
definition to be a weighted average of the spatially dependent
voltage $V_T(x,y,t)$,
\begin{equation}
V=\int \dif x \dif y u(x,y) V_T(x,y,t).
\label{eq:vaverage}
\end{equation}
This definition is chosen because it then follows from
Eq.~(\ref{eq:continuity}) that the product $IV$ gives the rate of
change of field energy in the cavity, and thus
Eq.~(\ref{eq:vaverage}) provides a reasonable definition of port
voltage. Solution of Eq.~(\ref{eq:virelation}) and application of
(\ref{eq:vaverage}) to the complex phasor amplitude $\hat{V}_T$
provide a linear relation between $\hat{V}$ and $\hat{I}$, which
defines the impedance $Z$.

To solve Eq.~(\ref{eq:virelation}), we  expand $\hat{V}_T$ in the
basis of the eigenfunctions of the closed cavity, i.e.,
$\hat{V}_T=\sum_n c_n \phi_n$, where $(\nabla ^2+k_n^2)\phi_n=0$,
$\int\phi_i\phi_j \dif x\dif y=\delta_{ij}$ and $\phi_n(x,y)=0$ at
the cavity boundary. Thus, multiplying Eq.~(\ref{eq:virelation}) by
$\phi_n$ and
integrating over $(x,y)$ yields
\begin{equation}
c_n(k^2-k_n^2)=-\mj kh\eta_0\langle u\phi_n\rangle\hat{I},
\end{equation}
where $k_n=\omega_n/c$, $\omega_n$ is the eigenfrequency
associated with $\phi_n$, and
$\langle\ u \phi_n \rangle=\int \phi_n u dxdy$.
 Solving for the
coefficients $c_n$ and computing the voltage $\hat{V}$ yields
~\begin{equation} \hat{V}=-\mj \sum_n \frac{kh\eta_0\langle u
\phi_n\rangle ^2} {k^2-k_n^2}\hat{I}=Z\hat{I}. \label{eq:zcav1}
\end{equation}
This equation describes the linear relation between the port
voltage and the current flowing into the port. Since we have
assumed no energy dissipation so far (e.g., due to wall absorption or
radiation), the impedance of the cavity is purely imaginary, as is
indicated by Eq.~(\ref{eq:zcav1}).

The expression for $Z$ in Eq.~(\ref{eq:zcav1}) is equivalent to a
formulation introduced by Wigner and Eisenbud \cite{Rtheory} in
nuclear-reaction theory in 1947, which was generalized and
reviewed by Lane and Thomas \cite{geneR1}, and Mahaux and
Weidenm\"{u}ller \cite{geneR2}. Recently,  a supersymmetry approach to
scattering based
on this formulation was introduced by Verbaarschot et.al. \cite{supersym1} 
and further developped by Lewenkopf \cite{supersym2} and Fyodorov
\cite{theo4}(which they called the
``$K$-matrix" formalism), and it has also been adapted to quantum dots by
Jalabert, Stone and Alhassid \cite{geneR3}.

Explicit evaluation of Eq.~(\ref{eq:zcav1}) in principle requires
determination of the eigenvalues and corresponding eigenfunctions
of the closed cavity. We do not propose to do this. Rather, we
adopt a statistical approach to replace $\langle u \phi_n \rangle$ and
$k_n^2$ with random variables with appropriate distribution, such that
we can construct models for the
statistical behavior of the impedance.
For high frequencies such that
$k=\omega/c\gg L^{-1}$ where $L$ is a typical dimension of the
cavity, the sum in Eq.~(\ref{eq:zcav1}) will be dominated by high
order (short wavelength) modes with $k_n L\gg 1$, and
the properties of the short wavelength eigenfunctions can be
understood in terms of ray trajectories. For geometries like that
in Fig.~\ref{fig:bowtie}(a), ray trajectories are chaotic.

The assumed form of the eigenfunction from the random plane wave
hypothesis is
\begin{equation}
\phi_n=\lim_{N\rightarrow
\infty}\sqrt{\frac{2}{AN}}Re\{\sum_{i=1}^{N}\alpha_i \exp(\mj k_n
\vec{e}_i\cdot \vec{x}+\mj\theta_i )\},
\label{eq:superposi}
\end{equation}
where $\vec e_i$ are randomly oriented unit vectors (in the
$x$-$y$ plane), $\theta _i$ is random in $[0,$ $2\pi]$, and
$\alpha _i$ are random.  This statistical model for $\phi_n$ is
motivated by the previously discussed ergodicity of ray paths in
chaotic cavities (e.g., Fig.~1(a)); i.e., the random orientation
of $\vec{e_i}$ corresponds to the uniform distribution of ray
orientations $\theta$.  Using (\ref{eq:superposi}) we can
calculate the overlap integral $\langle u\phi_n\rangle$ appearing
in the numerator of (\ref{eq:zcav1}). Being the sum of
contributions from a large number of random plane waves, the
central limit theorem implies that the overlap integral will be a
Gaussian random variable with zero mean. The variance of the
overlap integral can be obtained using Eq.~(\ref{eq:superposi}),
\begin{equation}
E\{\langle
u\phi_n\rangle^2\}=\frac{1}{A}\int_0^{2\pi}\frac{\dif\theta}{2\pi}
|\bar{u}(\vec{k}_n)|^2,
\label{eq:overlap}
\end{equation}
where $E\{ . \}$ denotes the expected value, $\bar{u}(\vec{k}_n)$ is
the Fourier transform of the profile function $u(x,y)$,
\begin{equation}
\bar{u}(\vec{k}_n)=\int \dif x \dif y u(x,y) exp(-\mj\vec{k}_n \cdot
\vec{x}),
\end{equation}
and $\vec{k}_n=(k_n \cos \theta , k_n \sin \theta )$. The integral
in (\ref{eq:overlap}) over $\theta$ represents averaging over the
directions $\vec{e_j}$ of the plane waves. The variance of $\langle u
\phi_n\rangle$ depends on the
eigenvalue $k^2_n$. If we consider a localized source $u(x,y)$
such that the size of the source is less than the typical
wavelength $2\pi/k_n$, then the variance will be $A^{-1}$ (recall
the normalization of $u$ given by Eq.~(\ref{eq:profileu})).

Modelling of Eq. (\ref{eq:zcav1}) also requires specifying the
distribution of eigenvalues $k_n$ appearing in the denominator.
According to the Weyl formula \cite{ott02}, for a two dimensional
cavity of area $A$, the average separation between adjacent
eigenvalues, $k_n^2-k_{n-1}^2$, is $4\pi A^{-1}$.  Thus, one
requirement on the sequence of eigenvalues is that they have a
mean spacing $4\pi A^{-1}$. The distribution of spacings of
adjacent eigenvalues is predicted to have the characteristic
Wigner form for cavities with chaotic trajectories. In particular,
defining the normalized spacing, $s_n=A(k_n^2-k_{n-1}^2)/4\pi $,
it is found that there are two basic cases which (for reasons
explained subsequently) are called ``time reversal symmetric"
(TRS) and ``time-reversal symmetry broken" (TRSB).  The
probability density function for $s_n$ is predicted to be closely
approximated by
\begin{equation}
P(s_n)=\frac{\pi}{2}s_n \exp(-\pi s_n^2/4)
\label{eq:spacetrs}
\end{equation}
for chaotic systems with time-reversal
symmetry(TRS) and
\begin{equation}
P(s_n)=\frac{32}{\pi}s_n^2 \exp(-4 s_n^2/\pi)
\label{eq:spacetrsb}
\end{equation}
for time-reversal symmetry broken(TRSB) system. Thus, a second
requirement on the sequence of eigenvalues is that they have the
correct spacing distribution.  The TRS case applies to systems
where the permittivity and permeability tensors are real and
diagonal. The TRSB case applies to systems where the  permittivity
or permeability tensors are complex but hermitian, as they are for
a magnetized ferrite.

One approach of ours will be to generate  values for the impedance
assuming that sequences of eigenvalues can be generated from a set
of separations $s_n$ which are independent and distributed
according to Eq.~(\ref{eq:spacetrs}). The usefulness of the
assumption of the independence of separations will have to be
tested, as it is known that there are long range correlations in
the spectrum, even if nearby eigenvalues appear to have
independent spacings. A more complete approach is to use a
sequence of eigenvalues taken from the spectra of random matrices.
When this is done the impedance defined in Eq.~(\ref{eq:zcav1})
(with independent Gaussian distributions for the overlap
integrals) is completely equivalent to that obtained in Random
Matrix Theory. We will find that in some cases it is sufficient to
consider the simpler spectra, generated from independent spacing
distributions, but in other cases, for example, when losses are
considered, or when correlations of impedance values at different
frequencies are considered, the correlations in eigenvalues
exhibited by random matrix theory are important. This will be
discussed more thoroughly later in the paper.

A key assumption in our model is the statistical independence of
the overlap integrals, $\langle u \phi_n \rangle$, and the
eigenvalues $k_n$. This we argue on the basis that each
eigenfunction satisfies the plane wave hypothesis and successive
eigenfunctions appear to be independent. A second justification
comes from random matrix theory where it is known that the
probability distribution for the eigenvalues of a random matrix is
independent of that of the elements of the eigenfunctions
(\cite{mehta91}, Chap. 3). Indeed, the result from the random
plane wave hypothesis (Eq.~(\ref{eq:zcav2}), below) turns out to
be equivalent to past work on scattering matrices that was based
on coupling to systems described by random matrix theory
\cite{mello85}.

Combining our expressions for $\langle u \phi_n \rangle$  and
using the result that for a two dimensional cavity the mean
spacing between adjacent eigenvalues is $\Delta =4\pi A^{-1}$, the
expression for the cavity impedance given in Eq.~(\ref{eq:zcav1})
can be rewritten,
\begin{equation}
Z=-\frac{\mj}{\pi}\sum_{n=1}^{\infty}\Delta
\frac{R_R(k_n)w_n^2}{k^2-k_n^2}, \label{eq:zcav2}
\end{equation}
where $w_n$ is taken to be a Gaussian random variable with zero
mean and unit variance, the $k_n$ are distributed independent of the
$w_n$, and $R_R$ is given by
\begin{equation}
R_R(k)=\frac{kh\eta_0}{4}\int\frac{\dif\theta}{2\pi}|u(\vec{k})|^2.
\label{eq:zr_kn}
\end{equation}

Our rationale for expressing the impedance in the form of
Eq.~(\ref{eq:zcav2}) and introducing $R_R(k_n)$ is motivated by
the following observation. Suppose we allow the lateral boundaries
of the cavity to be moved infinitely far from the port. That is,
we consider the port as a 2D free-space radiator. In this case, we
solve Eq. (\ref{eq:virelation}) with a boundary condition
corresponding to outgoing waves, which can be readily done by the
introduction of Fourier transforms. This allows us to compute the
phasor port voltage $\hat{V}$ by Eq.~(\ref{eq:vaverage}).
Introducing a complex radiation impedance $Z_R(k)=\hat{V}/\hat{I}$
(for the problem with the lateral boundaries removed), we have
\begin{equation}
Z_{R}(k)=-\frac{\mj}{\pi}\int_0^{\infty}\frac{\dif
k^2_n}{k^2-k_n^2}R_R(k_n), \label{eq:zfs}
\end{equation}
where $R_R(k_n)$ is given by Eq. (\ref{eq:zr_kn}) and $k_n$ is now
a continuous variable. The impedance $Z_{R}(k)$  is complex with a
real part obtained by deforming the $k_n$ integration contour to
pass above the pole at $k_n=k$. This follows as a consequence of
applying the outgoing wave boundary condition, or equivalently,
letting $k$ have a small negative imaginary part. Thus, we can
identify the quantity $R_{R}(k)$ in Eq.~(\ref{eq:zr_kn}) as the
radiation resistance of the port resulting from one half the
residue of the integral in (\ref{eq:zfs}) at the pole,
$k^2_n=k_n$,
\begin{equation}
Re[Z_R(k)]=R_R(k),
\label{eq:zfs_R}
\end{equation}
and
\[
X_{R}(k)=Im[Z_R(k)]
\]
 is the radiation reactance given by the
principal part (denoted by $P$) of the integral (\ref{eq:zfs}),
\begin{equation}
X_{R}(k)=P\{-\frac{1}{\pi}\int_0^{\infty}\frac{\dif
k^2_n}{k^2-k_n^2}R_R(k_n)\} .
\label{eq:zfs_I}
\end{equation}

Based on  the above, the connection between the cavity impedance,
represented by the sum in Eq.~(\ref{eq:zcav2}), and the radiation
impedance, represented  in Eq.~(\ref{eq:zfs_R}) and
Eq.~(\ref{eq:zfs_I}), is as follows. The cavity impedance,
Eq.~(\ref{eq:zcav2}), consists of a discrete sum over eigenvalues
$k_n$ with weighting coefficients $w_n$ which are Gaussian random
variables. There is an additional weighting factor $R_R(k_n)$ in
the sum, which is the radiation resistance. The radiation
reactance, Eq.~(\ref{eq:zfs_I}), has a form analogous to the
cavity impedance. It is the principle part of a continous integral
over $k_n$ with random coupling weights set to unity. While,
Eqs.~(\ref{eq:zcav2}), (\ref{eq:zfs_R}), (\ref{eq:zfs_I}), have
been obtained for the simple model input $\hat{J}=\hat{I} u(x,y)$
in $0\leq z \leq h$ with perfectly conducting plane surfaces at
$z=0, h$, we claim that these results apply in general. That is,
for a case like that in Fig.~1(b), $Z_R(k)$ (which for the simple
model is given by Eq.~(\ref{eq:zfs})) can be replaced by the
radiation impedance for the problem with the same port geometry.
We also note that while (\ref{eq:zfs}) was obtained with reference
to a two dimensional problem, the derivation and result are the
same in three dimensions. It is important to note that, while
$R_R(k)$ is nonuniversal (i.e., depends on the specific coupling
geometry, such as that in Fig. 2(b)), it is sometimes possible to
independently calculate it, and it is also a quantity that can be
directly measured (e.g., an experimental radiation condition can
be simulated by placing absorber  adjacent to the lateral walls).
In the next section, we will use the radiation impedance to
normalize the cavity impedance yielding a universal distribution
for the impedance of a chaotic cavity.

\section{Impedance  Statistics For a Lossless, Time Reversal Symmetric
Cavity}
In the lossless case, the impedance of the cavity $Z$ in
Eq.~(\ref{eq:zcav2}) is a purely imaginary number and $S$, the
reflection coefficient, is a complex number with unit modulus.
Terms in the summation of Eq.~(\ref{eq:zcav2}) for which $k^2$ is
close to $k_n^2$ will give rise to large fluctuations in $Z$ as
either $k^2$ is varied or as one considers different realizations
of the random numbers. The terms for which $k^2$ is far from
$k_n^2$ will contribute to a mean value of $Z$. Accordingly, we
write
\begin{equation}
Z=\bar{Z}+\tilde{Z},
\label{eq:zsep}
\end{equation}
where $\bar{Z}$, the mean value of $Z$, is written as
\begin{equation}
\bar{Z}=-\frac{\mj}{\pi}\sum_n \Delta
E\{\frac{R_R(k_n^2)}{k^2-k_n^2}\}, \label{eq:zbar1}
\end{equation}
and we have used the fact that the $w_n^2$ are independent with
$E\{w_n^2\}=1$. If we approximate the summation in
Eq.~(\ref{eq:zbar1}) by an integral, noting that $\Delta $ is the
mean spacing between eigenvalues, comparison with (\ref{eq:zfs_I})
yields
\begin{equation}
\bar{Z}=\mj X_{R}(k),
\label{eq:corresp1}
\end{equation}
where $X_R=Im[Z_R]$ is the radiation reactance defined by
Eq.~(\ref{eq:zfs_I}). Thus, the mean part of the fluctuating
impedance of a closed cavity is equal to the radiation reactance
that would be obtained under the  same coupling conditions for an
antenna radiating freely; i.e., in the absence of multiple
reflections of waves from the lateral boundaries of the cavity.
The equivalent conclusion for the radiation scattering coefficient
is evident from the treatment of Brouwer \cite{brouwer95}.

We now argue that, if $k^2$ is large enough that many terms in the
sum defining $Z$ satisfy $k_n^2<k^2$, then the fluctuating part of
the impedance $\tilde{Z}$ has a Lorentzian distribution with a
characteristic width $R_R(k)$. That is, we can write
\begin{equation}
Z=\mj(X_R+R_R\xi), \label{eq:corresp2}
\end{equation}
where $\xi$ is a zero mean unit width Lorentzian distributed
random variable, $P_{\xi}(\xi)=[\pi(1+\xi^2)]^{-1}$.

 Lorentzian distribution appears in the theory of nuclear scattering
\cite{kriega65} and arises as consequences of random matrix theory
\cite{mello95, theo4}. That the characteristic width scales as
$R_R(k)$ follows from the fact that the fluctuating part of the
impedance is dominated by terms for which $k_n^2\simeq k^2$. The
size of the contribution of a term in the sum in
Eq.~(\ref{eq:zcav2}) decreases as $|k^2-k^2_n|$ in the denominator
increases.  The many terms with large values of $|k^2-k^2_n|$
contribute mainly to the mean part of the reactance with the
fluctuations in these terms cancelling one another due to the
large number of such terms.  The contributions to the mean part
from the relatively fewer terms with small values of $|k^2-k^2_n|$
tend to cancel due to the sign change of the denominator while
their contribution to the fluctuating part of the reactance is
significant since there are a smaller number of these terms.
Consequently, when considering impedance fluctuations, it suffices
to  treat $R_R(k_n)$ as a constant in the summation in
Eq.~(\ref{eq:zcav2}) and factor it out. This results in a sum that
is independent of coupling geometry and is therefore expected to
have a universal distribution.

\subsection{Numerical Results for a Model Normalized Impedance}
To test the arguments above, we consider a model normalized cavity
reactance $\tilde{\xi}={X}/R_R$ and also introduce a normalized
wavenumber $\tilde k^2=k^2/\Delta =k^2A/4\pi $.  In terms of this
normalized wavenumber, the average of the eigenvalue spacing
[average of $(\tilde k^2_{n+1}-\tilde k^2_n)$] is unity.  Our
model normalized reactance is
\begin{equation}
\tilde{\xi}= -\frac{1}{\pi}\sum_{n=1}^{N}\frac{w_n^2}{\tilde
k^2-\tilde k_n^2}, \label{eq:x}
\end{equation}
where the $w_n$ are independent Gaussian random variables, $\tilde
k_n^2$ are chosen according to various distributions, and we have
set $R_R(k_n)$ to a constant value for $n\leq N$ and $R_R(k_n)=0$
for $n> N$. The fluctuating part of  $j\xi$ given by
Eq.~(\ref{eq:x}) mimics the fluctuating part of
the impedance $Z$ in the case in which $R_R(k_n)$ has a sharp
cut-off for eigenmodes with $n>N$. In terms of $\xi$,
Eq.~(\ref{eq:corresp2}) becomes
\begin{equation}
P_{\tilde{\xi}}(\tilde{\xi})=\frac{1}{\pi}\frac{1}{[(\tilde{\xi}-\bar
\xi )^2+1]}, \label{eq:pdfxi}
\end{equation}
where $\bar \xi $ is the mean of $\xi $.

 First we consider the
hypothetical case where the collection of $\tilde k_n^2$ values
used in Eq.~(\ref{eq:x}) result from $N$ independent and uniformly
distributed random choices in the interval $0 \leqslant \tilde
k_n^2\leqslant N$. In contrast to Eqs.~ (\ref{eq:spacetrs}),  this
corresponds to a Poisson distribution
of spacings $P(s)=exp(-s)$ for large $N$. 
This case is analytically solvable \cite{mello95} and that the mean value
$\bar \xi $ is
\begin{equation}
\bar{\xi}=P\{-\frac{1}{\pi}\int_{0}^{N}\frac{d\tilde k_n^2}{\tilde
k^2-\tilde k_n^2}\} =\frac{1}{\pi}ln|\frac{N-\tilde k^2}{\tilde
k^2}|, \label{eq:xmean}
\end{equation}
and, furthermore, that  $\xi $ has a Lorentzian distribution given by
Eq.~(\ref{eq:pdfxi}).

\begin{figure}
\includegraphics[scale=0.45]{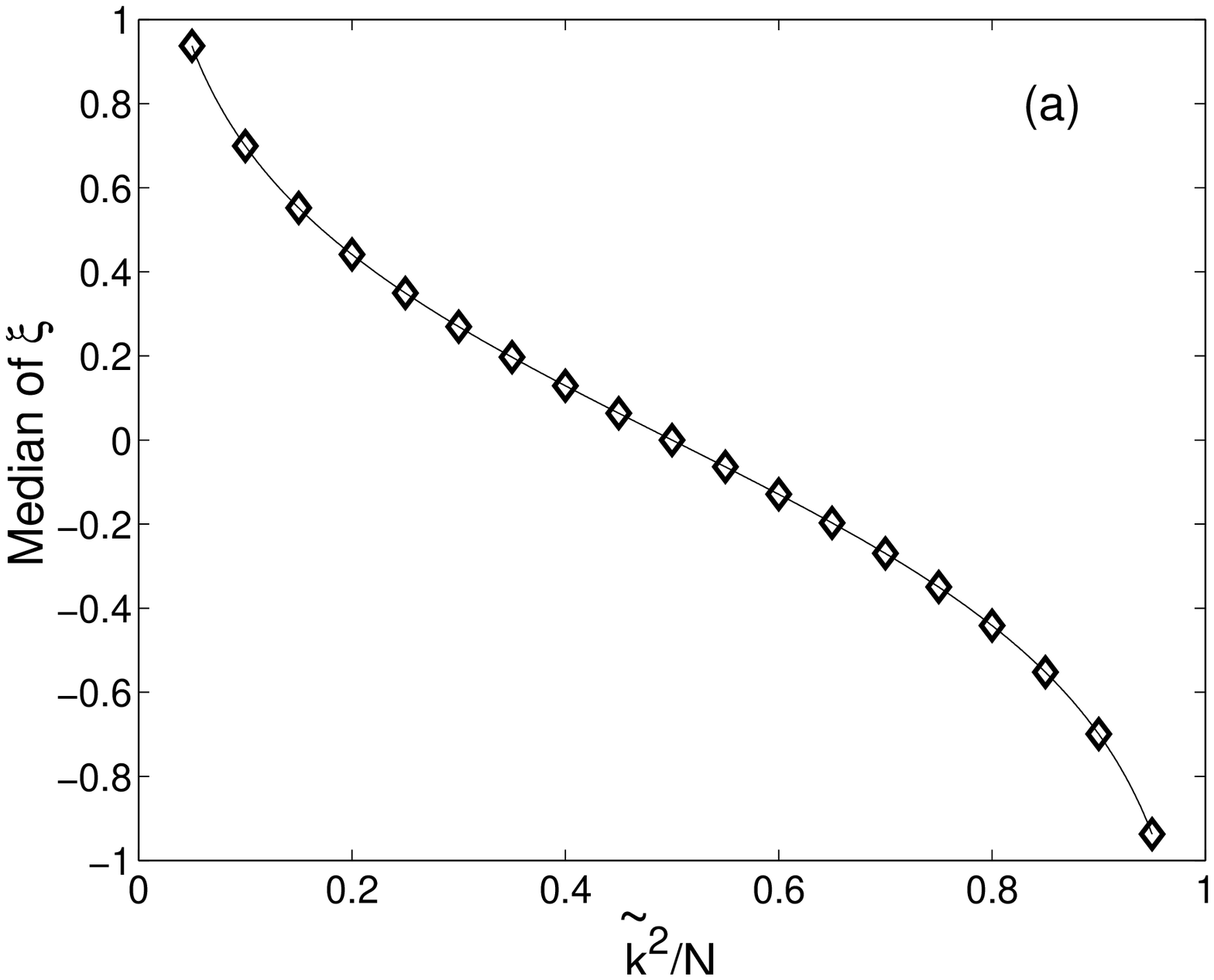}
\includegraphics[scale=0.45]{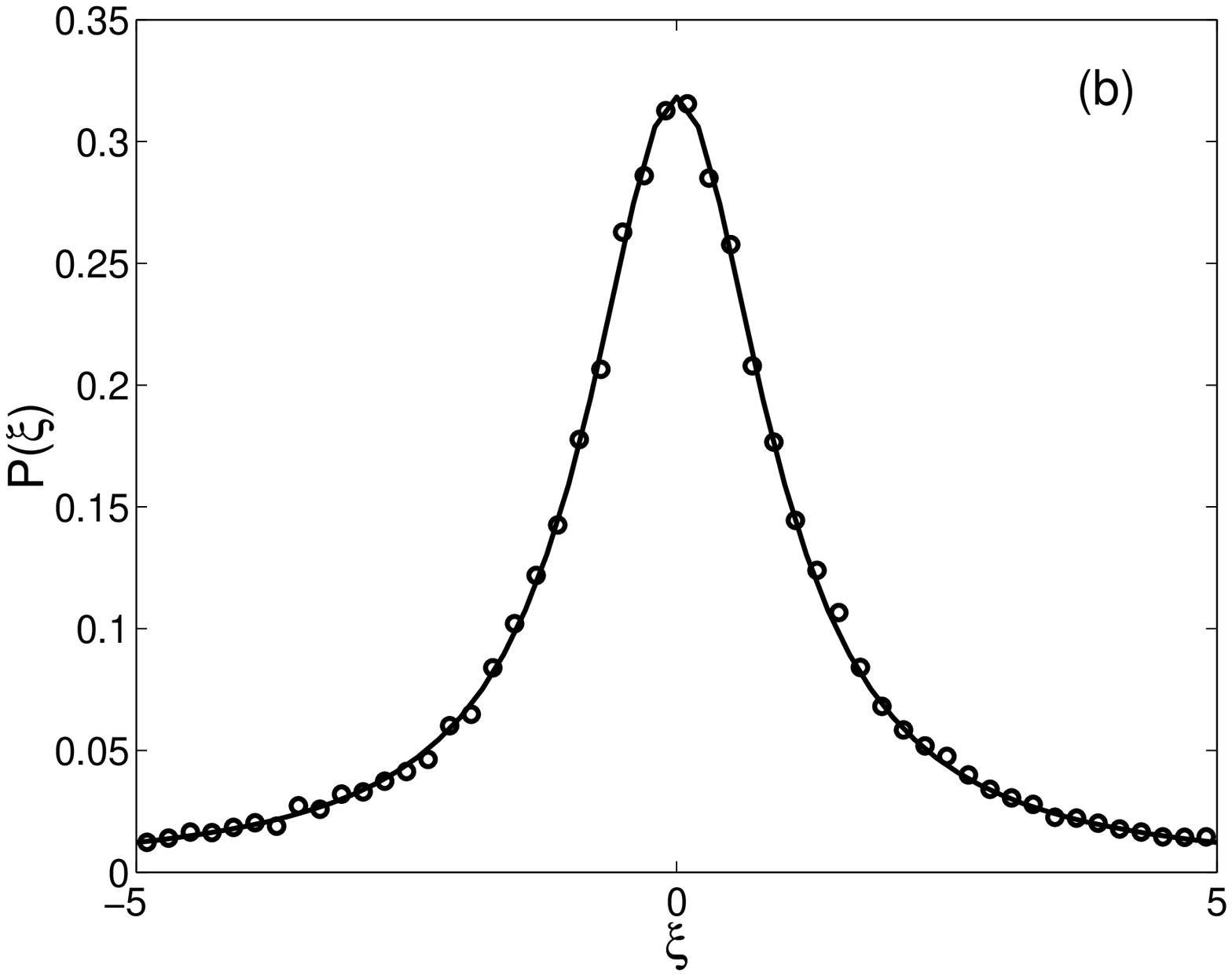}
\includegraphics[scale=0.45]{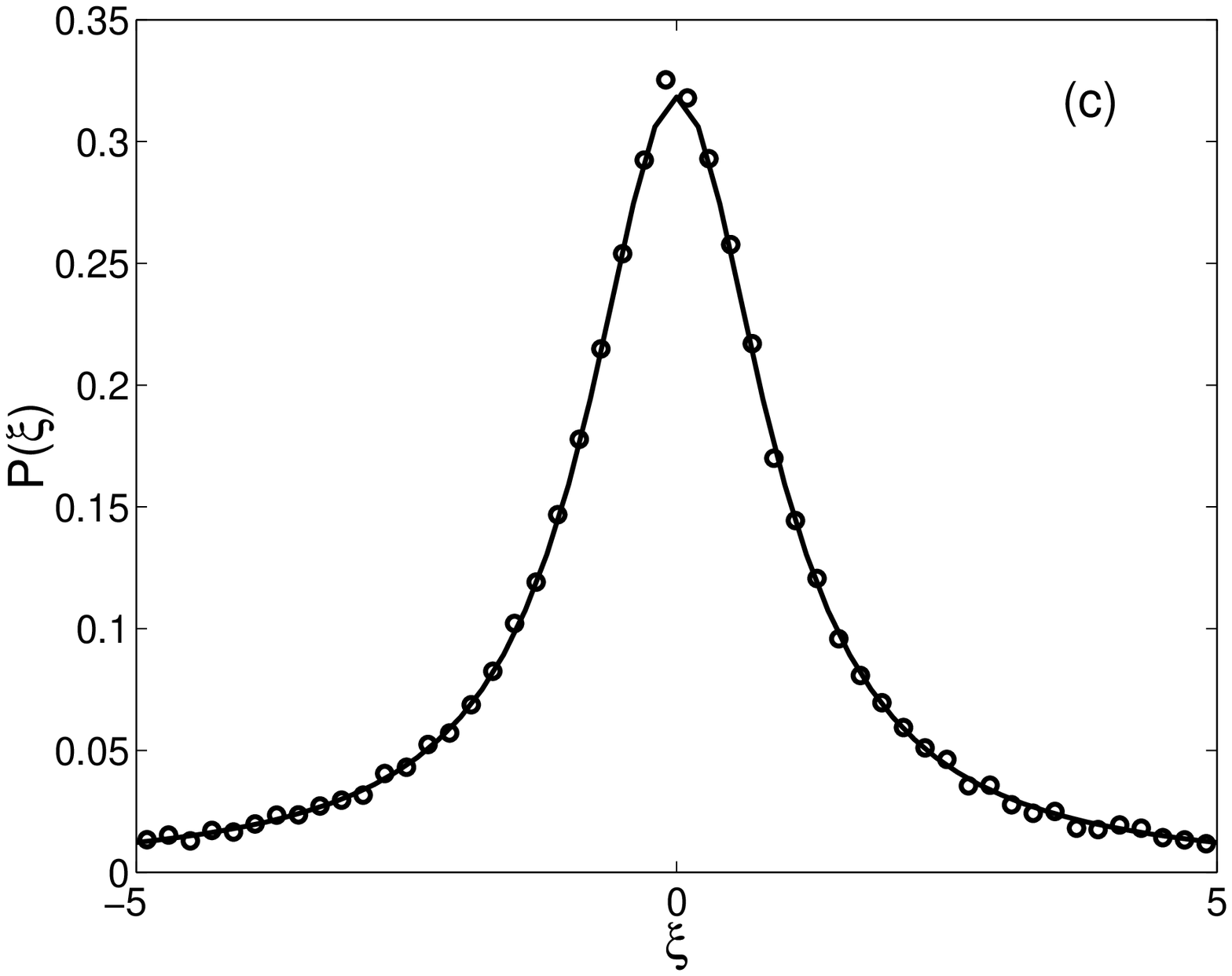}
\caption{ (a) Median of $\xi$ versus  $\tilde k^2/N$, compared
with Eq.~(\ref{eq:xmean}). (b) Histogram of approximation to
$P_{\xi}(\xi)$ (solid dots) in the TRS case compared with a
Lorentzian distribution of  unit width. (c) Same as (b) but for
the  TRSB case. } \label{fig:xi}
\end{figure}

Our next step is to numerically determine the probability distribution
function for $\xi$ given by (\ref{eq:x}) in the case where the
spacing distribution corresponds to the TRS case described by
Eq.~(3). We generated $10^6$ realizations of the
sum in Eq.~(\ref{eq:x}). For each realization we randomly
generated $N=2000$ eigenvalues using the spacing probability
distribution (3), as well as $N=2000$ random values of $w_n$
chosen using a Gaussian distribution for $w_n$ with $E\{ w_n\} =0$
and $E\{ w_n^2\} =1$. We first test the prediction of
Eq.~(\ref{eq:xmean}) by plotting the median value of $\xi$ versus
$\tilde k^2$ in Fig.~\ref{fig:xi}(a). (We use the median rather
than the mean, since, for a random variable with a Lorentzian
distribution, this quantity is more robust when a finite sample
size is considered.) Also plotted in Fig.~\ref{fig:xi}(a) is the
formula (\ref{eq:xmean}). We see that the agreement is very good.
Next we test the prediction for the fluctuations in $\xi$ by
plotting a histogram of $\xi$ values for the case $\tilde k^2=N/2$
in Fig.~\ref{fig:xi}(b). From (\ref{eq:xmean}) for $\tilde
k^2=N/2$ the mean is expected to be zero, and, as can be seen in
the figure, the histogram (open circles) corresponds to a Lorentzian with
zero mean and unit width (solid line) as expected. Histograms plotted for
other values of $\tilde k^2$ agree with the prediction but are not
shown. Thus, we find that the statistics of $\xi$ are the same for
$P(s)=exp(-s)$ (Poisson) and for $P(s)$ given by
Eq.~(\ref{eq:spacetrs}). Hence we conclude that the statistics of
$\xi$ are independent of the distribution of spacings. This is
further supported by Fig.~\ref{fig:xi}(c) where the histogram of
$\xi$ for $\tilde k^2=N/2$ is plotted for the case in which the
spacing distribution is that corresponding to time reversal
symmetry broken (TRSB) systems. (the TSRB
case will be discussed more carefully in a subsequent paper). Again
the histogram is in excellent agreement with (\ref{eq:pdfxi}).
This implies that, for the lossless case, with a single input transmission
line to the cavity, the impedance statistics are not so sensitive to
the spacing distributions, as long as they have the same mean value.

The issue of long range correlations in the
distribution of eigenvalues seems doesn't affect  statistics of the
impedance in the lossless case. In principle, one can  also
incorporate additional eigenvalue correlation from random matrix theory
in the statistics generating the $k_n^2$ in Eq.~(\ref{eq:x}).(and when
losses are considered, this is necessary.) We note
that the mean and width of the distribution in
the random matrix approach are specific to the random matrix problem. In
contrast, in our formulation, these quantities are determined by
the geometry specific port coupling to the cavity through the
radiation impedance $Z_R(k_n^2)$.

\subsection{HFSS simulation result for the normalized impedance}
\begin{figure}
\includegraphics[scale=0.4]{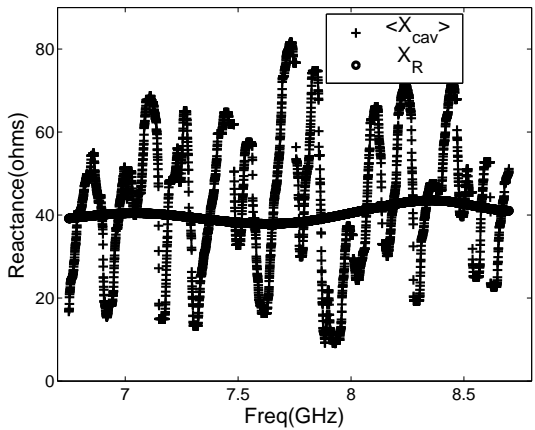}
\caption{Median cavity reactance averaged over 100 realization
vs. frequencies ranged from 6.75GHz to 8.75GHz, compared with
the corresponding radiation reactance $Im[Z_R(\omega)]$.}
\label{fig:medreact}
\end{figure}

To test our prediction for the distribution function of the
normalized impedance,  we have computed the impedance for the
cavity in Fig.~\ref{fig:bowtie}(a) for the coupling shown in
Fig.~\ref{fig:bowtie}(b) using the commercially available program
HFSS (High Frequency Structure Simulator \cite{hfss}). To create
different realizations of the configuration, we placed a small
metallic cylinder of radius 0.6 cm  and height $h$ at 100
different points inside the cavity. In addition, for each location
of the cylinder, we swept the frequency through a 2.0 GHz range
(about 100 modes) from 6.75GHz to 8.75GHz in 4000 steps of width
$5\times10^{-4}$ GHz. We generated 100,000 impedance values. In
addition, to obtain the radiation impedance, we also used HFSS to
simulate the case with radiation  boundary conditions assigned to
the sidewalls of the cavity.  We find that the average value of
the cavity reactance (which we predict to be the radiation
reactance) has large systematic fluctuations. This is illustrated
in Fig.~\ref{fig:medreact} where we plot the median cavity
reactance versus frequency. Here the median is taken with respect
to the 100 locations of the perturbing disc. Also shown in
Fig.~\ref{fig:medreact} is the radiation reactance
$X_R(\omega)=Im[Z_R(\omega)]$. As can be seen the radiation
reactance varies only slightly over the plotted frequency range,
whereas the median cavity reactance has large frequency dependent
fluctuations about this value. On the other hand, we note that
over the range 6.75-8.75 GHz, the average radiation reactance is
40.4 $\Omega$ and the average of the median cavity reactances is
42.3$\Omega$. Thus over this frequency band, there is good
agreement. The scale of the fluctuations in cavity reactance is on
the order of 0.2GHz, which is much larger than the average spacing
between cavity resonances which is only 0.016GHz. Thus, these
fluctuations are not associated with individual resonances.
Rather, the frequency scale of 0.2GHz suggests that they are
multipath interference effects ($L\sim 100cm$), which survive in
the presence of the moveable conducting disc. One possibility is
that the fluctuations are the result of scars \cite{heller84} and
this will be investigated in the future. The implication of
Fig.~\ref{fig:medreact} is that to obtain good agreement with the
theory predicting a Lorentzian distribution, it may be necessary
to average over a sufficiently large frequency interval.

\begin{figure}
\includegraphics[scale=0.4]{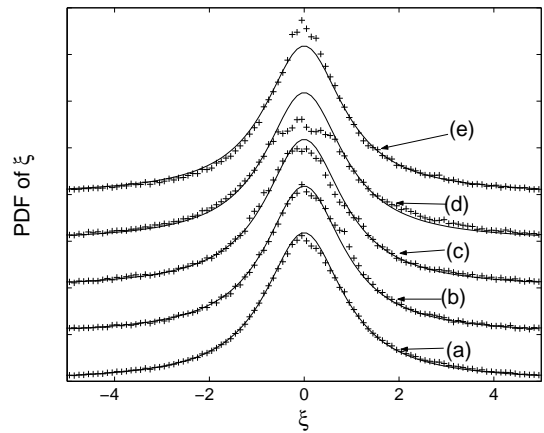}
\caption{Histogram approximation to $P_{\xi}(\xi)$ from numerical
data calculated using HFSS in different frequency ranges. (a) 6.75 - 8.75
GHz, (b) 6.75 - 7.25 GHz, (c) 7.25 - 7.75GHz, (d) 7.75 - 8.25 GHz,
(e) 8.25 - 8.75 GHz.}
\label{fig:hfssX}
\end{figure}

To test the Lorentzian prediction we normalize the cavity
impedance using the radiation impedance as in
Eq.~(\ref{eq:corresp1}) and Eq.~(\ref{eq:corresp2}), the
normalized impedance values, $\tilde{\xi}= \{ {\rm
Im}[Z(k)]-X_R(k)]\}/R_R(k)$, are computed, and the resulting
histogram approximations to $P_{\xi}(\tilde{\xi})$ is obtained.
Fig.~\ref{fig:hfssX}(a) shows the result for the case where we
have used data in the frequency range 6.75GHz to 8.75GHz (the
range plotted in Fig.~\ref{fig:medreact}). The histogram points
are shown as dots, and the theorectical unit width Lorentzian is
shown as a solid curve. Good agreement between the predicted
Lorentzian and the data is seen. Figures \ref{fig:hfssX} (b)-(e)
show similar plots obtained for smaller frequency range of width
0.5GHz: (b) 6.75 - 7.25 GHz, (c) 7.25 - 7.75GHz, (d) 7.75 - 8.25
GHz, (e) 8.25 - 8.75 GHz. For these narrow freqency ranges, we see
that Figs.~\ref{fig:hfssX}(b) and \ref{fig:hfssX}(c) show good
agreement with (\ref{eq:pdfxi}), while, on the other hand,
Figs.~\ref{fig:hfssX}(d) and \ref{fig:hfssX}(e)  exhibit some
differences. These are possibly associated with the variances in
the median cavity reactance shown in Fig.~3 as the agreement with
the Lorentzian prediction improves when averaging over a large
range of frequencies.

\subsection{Variation in Coupling}
\begin{figure}
\includegraphics[scale=0.4]{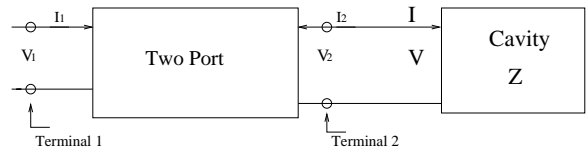}
\caption{Schematic description of the two port model}
\label{fig:twoport}
\end{figure}

In this section, we bolster our arguments connecting the radiation impedance  and
the normalization
of the cavity impedance by showing that the relation is preserved when the details
of
the coupling port are modified.
Let us consider a one-port coupling case in
which the actual coupling is equivalent to the cascade of a
lossless two port and a ``pre-impedance" $Z$ seen at terminal 2,
as illustrated in
Fig.~\ref{fig:twoport}.

The impedance $Z$ at terminal 2 then transforms to a new
impedance $Z'$
at terminal 1 of the two port according to
\begin{equation}
Z'=\mj \hat{X}_{11}+\frac{\hat{X}_{12}\hat{X}_{21}}{\mj\hat{X}_{22}+Z},
\label{eq:zeq}
\end{equation}
where $j\hat{X}_{ij}$ is now the purely imaginary 2 by 2 impedance
matrix of the lossless two-port.  We now ask how $Z$ transforms to
$Z'$ when (a) $Z$ is the complex impedance $Z_R$ corresponding to
the radiation impedance into the cavity (i.e. the cavity
boundaries are extended to infinity) and (b) $Z=\mj X$ is an
imaginary impedance corresponding to a lossless cavity, where
 $X$ has a mean $\bar{X}$ and Lorentzian distributed fluctuation
$\tilde{X}$.

First considering case (a) the complex cavity impedance
$Z_R=R_{R}+\mj X_{R}$ transforms to a complex impedance
$Z'_R=R'_{R}+ \mj X'_{R}$ where
\begin{equation}
R'_{R}=R_{R}\frac{\hat{X}_{12}\hat{X}_{21}}{R_{R}^2+(\hat{X}_{22}+X_R)^2},
\end{equation}
and
\begin{equation}
X'_R=\hat{X}_{11}-(\hat{X}_{22}+X_R)\frac{\hat{X}_{12}\hat{X}_{21}}
{R_R^2+(\hat{X}_{22}+X_R)^2}.
\end{equation}
In case (b) we consider the transformation of the random variable
$X$ to a new random variable $X'$ according to
$X'=\hat{X}_{11}+\hat{X}_{12}\hat{X}_{21}/(\hat{X}_{22}+X)$. One
can show that if $X$ is Lorentzian distributed with mean $X_R$ and
width $R_R$ then $X'$ will be Lorentzian distributed with mean
$X'_R$ and the width $R'_{R}$. Thus, the relation between the
radiation impedance and the fluctuating cavity impedance is
preserved by the lossless two port. Accordingly, we reassert that
this relation holds in general for coupling structures whose
properties are not affected by the distant walls of the cavity. A
treatment similar to that above has also been given by Brouwer
\cite{brouwer95} in the context of scattering with a scattering
matrix description of the connection between terminal 1 and 2.

We now summarize the main ideas of this section. The normalized
impedance of a lossless chaotic cavity with time-reversal symmetry has a
universal distribution which is a Lorentzian. The width of the
Lorentzian and the mean value of the impedance can be obtained by
measuring the corresponding radiation impedance under the same
coupling conditions. The physical interpretation of this
correspondence is as follows. In the radiation impedance, the
imaginary part is determined by the near field, which is
independent of cavity boundaries.  On the other hand, the real
part of the radiation impedance is related to the far field. In a
closed, lossless cavity, the real part of the impedance vanishes.
However, waves that are radiated into the cavity are reflected
from the boundaries eventually returning to the port and giving
rise to fluctuation in the cavity reactance.

\section{Statistics of Reflection Coefficient in the Lossless Case}
In the previous section, we obtained a universal Lorentzian
distribution for the chaotic cavity impedance $Z$, after
normalization by the radiation impedance,
\begin{equation}
Z=j(X_R+R_R{\xi}), \label{eq:zfinal}
\end{equation}
where ${\xi}$ is a zero mean, unit width Lorentzian random
variable. We now  consider the consequences for the reflection
coefficient. Suppose we can realize the perfect coupling
condition, i.e. $R_R=Z_0$, $X_R=0$, in which the wave does not
``feel"   the transition from the cable to the cavity. In this
case the cavity reflection coefficient becomes
\begin{equation}
S=\frac{j{\xi}-1}{j{\xi}+1}=\exp[-j(2\tan^{-1}{\xi}+\pi)].
\label{eq:spc}
\end{equation}
A standard Lorentzian distribution for ${\xi}$  corresponds to a
uniform distribution for $\tan^{-1}{\xi}$ from [$-\pi/2$,
$\pi/2$], and thus to a reflection coefficient uniformly
distributed on the unit circle.

In the general case (i.e., non-perfect coupling), we
introduce $\gamma_R=R_R/Z_0$, $\gamma_X=X_R/Z_0$, and express $S$ as
\begin{equation}
S=\me ^{\mj\phi}=(Z+Z_0)^{-1}(Z-Z_0)= \frac{\mj(\gamma_R
\xi+\gamma_X)-1} {\mj(\gamma_R \xi+\gamma_X)+1}. \label{eq:s}
\end{equation}
 We replace the
Lorentzian random variable $\xi$ by introducing another
random variable $\psi$ via $\xi=\tan(\psi/2)$. Using this
substitution, the Lorentzian distribution of $ \xi$ translates to
a distribution of $\psi$ that is uniform in [0, $2\pi$]. We then
have from Eq.~(\ref{eq:s})
\begin{equation}
\me^{\mj(\phi-\phi_R)}=\frac{\me^{-\mj\psi
'}+|\rho_R|}{1+|\rho_R|\me^{-\mj\psi'}},
\label{eq:phimphir}
\end{equation}
where the ``free space reflection coefficient" $\rho_R$
\begin{equation}
\rho_R=|\rho_R|\me^{\mj\phi_R}=\frac{\gamma_R+\mj\gamma_X-1}
{\gamma_R+\mj\gamma_X+1},
\label{eq:rhor}
\end{equation}
is the complex reflection coefficient in the case in which the
cavity impedance is set equal to the radiation impedance ($\tilde
\xi=-\mj$), and $\psi
'=\psi+\pi+\phi_R+2\tan^{-1}[\gamma_X/(\gamma_R+1)]$ is a shifted
version of $\psi$. Equations for the magnitude and phase of the
free space reflection coefficient $\rho _R$ can be obtained from
Eq.~(\ref{eq:rhor}). Specifically,
\begin{equation}
|\rho_R|=\sqrt{\frac{(\gamma_R-1)^2+\gamma_X^2}{(\gamma_R+1)^2+\gamma_X^2}},
\label{eq:magrhoR}
\end{equation}
and
\begin{equation}
\tan\phi_R=\frac{2\gamma_X}{\gamma_R^2+\gamma_X^2-1}.
\label{eq:angrhoR}
\end{equation}
Eq.~(\ref{eq:phimphir}) is essentially a statement of the Poisson
kernel relation for a non-perfectly coupled one port cavity.

To compute the probability distribution function for $\phi$, $
P_{\phi}(\phi)$, we note that, since $\psi$ is uniformly
distributed on any interval of $2\pi$, we can just as well take
$\psi '$, which differs from $\psi$ by a constant shift, to be
uniformly distributed. Consequently, we have
\begin{equation}
\begin{aligned}
P_{\phi}(\phi)&=\frac{1}{2\pi}|\frac{d\psi '}{d\phi}|\\
&=\frac{1}{2\pi}\frac{1-|\rho_R|^2}{1+|\rho_R|^2-2|\rho_R|\cos(\phi-\phi_R)}.
\end{aligned}
\label{eq:phasepdf}
\end{equation}
Thus $P_{\phi}(\phi)$ is peaked at the angle $\phi_R$
corresponding to the phase angle of the free space reflection
coefficient, with a degree of peaking that depends on $|\rho_R|$, the
magnitude of the free space reflection coefficient. `Perfect
matching' corresponds to $\gamma_R=1$, $\gamma_X=0$, and
$|\rho_R|=0$, in which case $P_{\phi}(\phi)$ is uniform.

We next consider the case of poor matching for which
$|\rho_R|\cong 1$ and $P_{\phi}(\phi)$ is strongly peaked at
$\phi_R$. This behavior can be understood in the context of the
frequency dependence of the phase for a given realization. It
follows from (\ref{eq:s}) and (\ref{eq:x}) that the phase $\phi$
decreases by $2\pi$ as $k^2$ increases by the spacing between
eigenvalues. If $|\rho_R| \cong 1$, then for most of the
frequencies in this interval, the phase is near $\phi_{R}$.
However, for the small range of frequencies near a resonance, the
phase will jump by $2\pi$ as the resonance is passed. This
indicates that the mode of the cavity is poorly coupled to the
transmission line. In the case of good matching, $|\rho_R|=0$, all
phases are equally likely indicating that, as a function of
frequency, the rate of increase of phase is roughly constant. This
implies that the resonances are broad, and the cavity is well
coupled to the transmission line.

In order to describe the different coupling strengths, we consider
the parameter $g$ originally introduced by  Fyodorov and Sommers
\cite{theo4} :
\begin{equation}
g=\frac{1+|\langle \me^{\mj\phi} \rangle |^2}{1-|\langle
\me^{\mj\phi} \rangle|^2}.
\label{eq:before_Gdef}
\end{equation}
Evaluating $\langle S \rangle$ using Eq.~(\ref{eq:phasepdf}),
\begin{equation}
g=\frac{1+|\rho_R|^2}{1-|\rho_R|^2}.
\label{eq:Gdef}
\end{equation}
Thus, $g$ has a minimum value of 1 in the perfectly matched case
and is large if the matching is poor, $|\rho_R|\sim 1$. An
analogous quantity is the voltage standing wave ratio on the
transmission line when the cavity impedance is set equal to the
radiation impedance,
\begin{equation}
{\rm VSWR}=\frac{1+|\rho_R|}{1-|\rho_R|}=g+\sqrt{g^2-1}.
\end{equation}

\begin{figure}
\includegraphics[scale=0.3,angle=270]{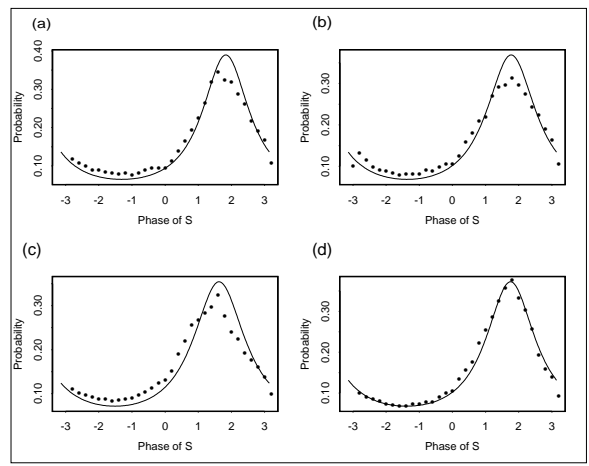}
\caption{Histogram of the reflection phase distribution for an
HFSS calculation for the cavity in Fig.~\ref{fig:bowtie} with
center frequencies located at (a) 7GHz, (b) 7.5GHz, (c) 8GHz, (d)
8.5GHz, and with sweeping span equal to 0.1GHz. Numerical data are
compared with Eq.~(\ref{eq:phasepdf}) using parameters determined
by $Z_{R}$ at the corresponding center frequencies.}
\label{fig:histS}
\end{figure}

To test Eq.~(\ref{eq:phasepdf}), we compared its predictions for
the phase distribution with direct numerical calculations obtained
using HFSS (High Frequency Structure Simulator) for the case of
the cavity and coupling detail as specified in
Fig.~\ref{fig:hfssX}. As compared to what was done for
Fig.~\ref{fig:hfssX}, we have narrowed the frequency range to 0.1
GHz bands for each realization in 1000 $10^{-4}$ GHz steps
centered at 7 GHz, 7.5 GHz, 8 GHz, 8.5 GHz. Instead of calculating
the radiation impedance for every frequency, we use the value of
$Z_{R}$ at the middle frequency of the interval in calculating the
values of $\gamma_R$ and $\gamma_X$. We present theoretical phase
density distribution functions together with numerical histogram
results in Fig.~\ref{fig:histS}.  The agreement between the
theory, Eq.~(\ref{eq:phasepdf}), and the numerical results is
surprisingly good, especially considering the rather small
(0.1GHz) frequency range used.

\section{Effect of distributed losses}
We now consider the effect of distributed losses in the cavity. By
distributed losses, we mean losses that affect all modes in a
frequency band equally (or at least approximately so). For
example, wall losses and losses from a lossy dielectric that fills
the cavity are considered distributed. [For the case of losses due
to conducting walls, the losses are approximately proportional to
the surface resistivity, $\sim \sqrt{f }$, and vary little in a
frequency range $\Delta f \ll f$. In addition, there will also be
variation of wall losses from mode to mode due to different
eigenmode structural details. These modal fluctuations, however,
are small when the modes are chaotic and the wavelength is short.]
We use the random coupling model to construct a complex
cavity impedance  accounting for distributed losses in a manner
analogous to the lossless
case, Eq.~(\ref{eq:zcav2}),
\begin{equation}
Z=-\frac{j}{\pi}\sum_n \Delta \frac{R_R(k_n)  w_n^2 }{k^2(1-j\sigma)-k_n^2},
\label{eq:zfinal_loss}
\end{equation}
where $\sigma$ represents the effect of losses. In particular, for
loss due to  wall absorption in a two-dimensional cavity, the
value of $\sigma$ is equal to the ratio of the skin depth of the
conductor to the height of the cavity; if the cavity contains a
lossy dielectric, $\sigma$ is the loss tangent of the dielectric.
The cavity quality factor is related to $\sigma$ by
$\sigma=Q^{-1}$. This follows by noting that the real part of $Z$
will have a Lorentzian dependence on frequency ($\omega=kc$)
peaking at $\omega=k_nc$ with a full width at half maximum of
$\omega \sigma$.

The impedance $Z$ will have a real part and an imaginary part. We expect that, if
$k^2\sigma\ll \Delta$, corresponding
to small losses, then  the real part will be zero and the imaginary part
will have
an approximately Lorentzian distribution. As losses are
increased  such that $k^2\sigma \sim \Delta$ (the imaginary part of the
denominators in (\ref{eq:zfinal_loss}) is of the order of
eigenvalue spacing), the distributions of the real and imaginary part
will change,
reflecting that extremely large values of $|Z|$ are
no longer likely. In the high loss limit, $k^2\sigma \gg\Delta$,  many terms in
the sum  contribute to the value of $Z$. In this case,
we expect $Z$ will approach the radiation impedance with small (Gaussian) fluctuations.

In the Appendix we evaluate the mean and variance of the real and
imaginary part of
the complex impedance (\ref{eq:zfinal_loss})
$Z=R+jX$.  There it is shown that the mean is the radiation impedance
$Z_R=R_R+jX_R$, and the variances of the real and imaginary
parts are equal $Var[R]=Var[X]$. In general, the distribution of $R$ and $X$ depends
on the correlations between eigenvalues of
$k_n^2$. However, in the low damping limit, the correlations are unimportant and we obtain
\begin{equation}
Var[R]=\frac{3R_R^2}{2\pi}\frac{\Delta}{k^2\sigma}
\label{eq:var_lodamp}\end{equation}
for both the TRS and the TRSB cases.
In the high damping limit $k^2\sigma\gg \Delta$, correlations are important and we obtain
\begin{equation}
\begin{aligned}
Var[R]&=\frac{R_R^2}{\pi}\frac{\Delta}{k^2\sigma} \qquad \text{for the TRS case}\\
Var[R]&=\frac{R_R^2}{2\pi}\frac{\Delta}{k^2\sigma} \qquad \text{for the TRSB
case.}
\end{aligned}
\label{eq:varR2case}
\end{equation}
This is to be constrasted with the result one would obtain if correlations in the
eigenvalue spacing were neglected; i.e., if the $k_n$ were assumed to be generated
by adding independent spacings generated from the distributions
(\ref{eq:spacetrs}) and (\ref{eq:spacetrsb}). In that case, using the method in
the Appendix  one obtains
\begin{equation}
\begin{aligned}
Var[R]&=\frac{R_R^2}{\pi}\frac{\Delta}{k^2\sigma}(\frac{1}{2}+\frac{2}{\pi}) \qquad
\text{for the TRS case}\\
Var[R]&=\frac{R_R^2}{\pi}\frac{\Delta}{k^2\sigma}(\frac{3\pi}{16}) \qquad
\text{for the TRSB
case.}
\end{aligned}
\label{eq:var_indep}
\end{equation}
These results are larger than those in Eq.~(\ref{eq:varR2case}) by $13.7\%$
in
the TRS case and $17.8\%$ in the TRSB case, thus illustrating the necessity of
generating the $k_n^2$ using random matrix theory if accurate results are desired
in the lossy case $k^2\sigma > \Delta$.

In a recent experimental paper \cite{warne03} the impedance statistics of
a lossy TRS one-port
microwave cavity were also considered. Their result is the same as
(\ref{eq:zfinal_loss}). One difference is that they generate the realizations of
$k_n^2$ solely by use of Eq.~(\ref{eq:spacetrs}) with the assumption that the
eigenvalue spacings are random independent variables.

We now investigate a model, normalized impedance, applicable in the
one-port case with loss, which is the generalization of Eq.~(\ref{eq:x}),
\begin{equation}
\zeta(\sigma)=-\frac{\mj}{\pi}\sum_{n=1}^{N}\frac{w_n^2}{\tilde
k^2(1-j\sigma)-\tilde k_n^2}. \label{eq:xsig}
\end{equation}
The normalized impedance $\zeta$
will have a real part $\rho >0$ and an imaginary part $\xi$,
$\zeta =\rho +j\xi $. We expect that if $\tilde k^2\sigma \ll 1$,
corresponding to small loss, then $\rho \cong 0$, and $\xi$ will have
an approximately Lorentzian distribution.

\begin{figure}
\includegraphics[scale=0.4]{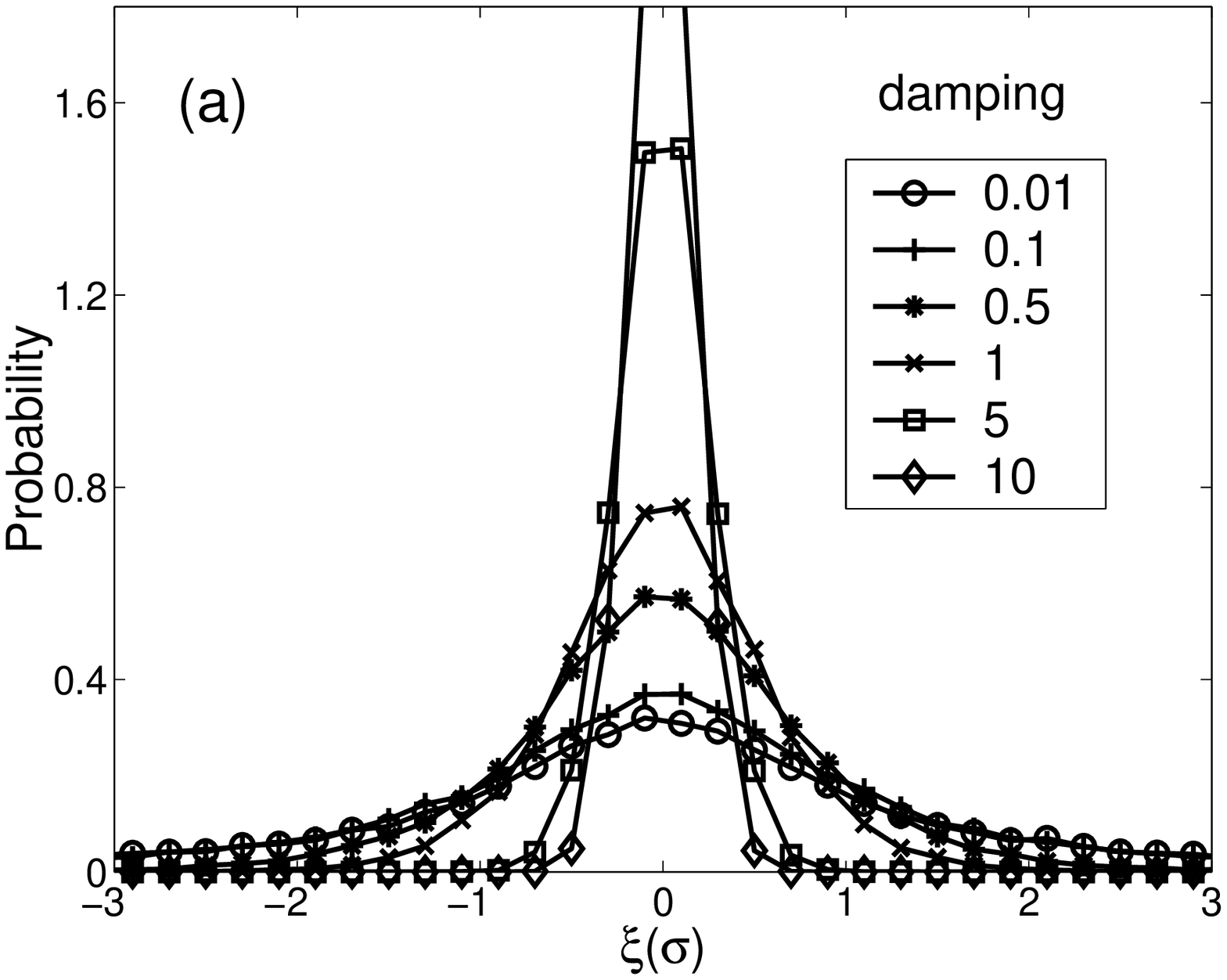}
\includegraphics[scale=0.4]{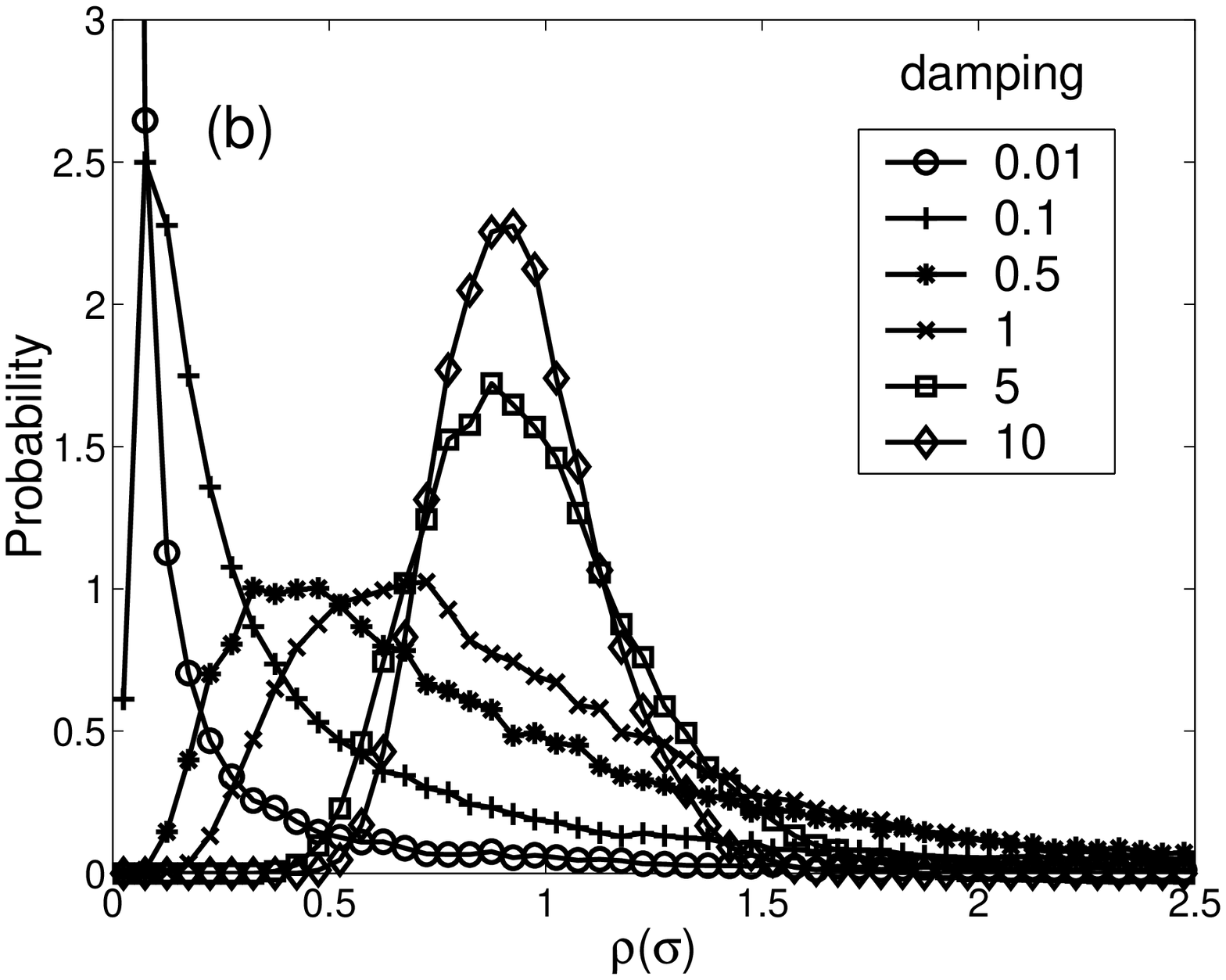}
\includegraphics[scale=0.4]{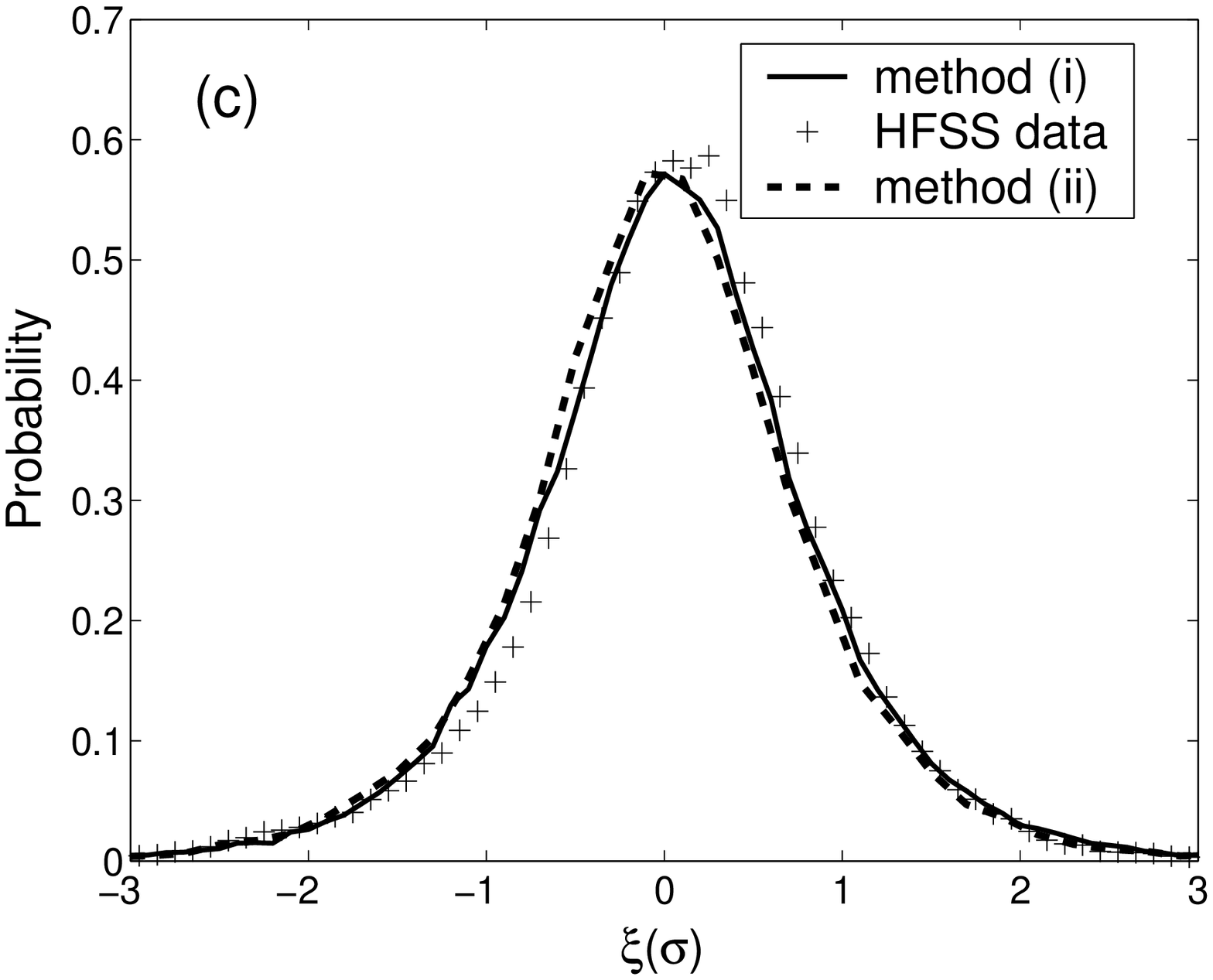}
\includegraphics[scale=0.4]{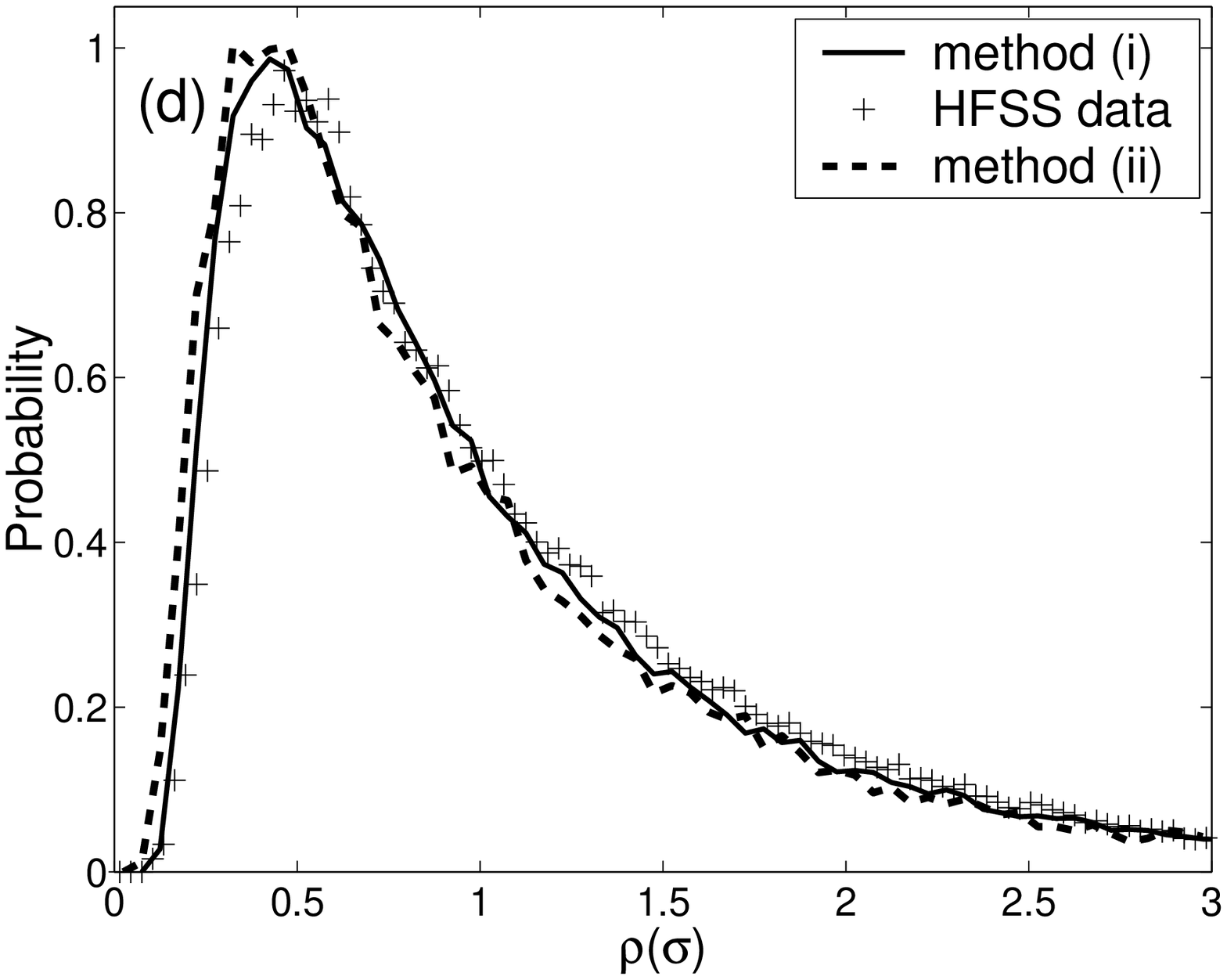}
\caption{(a) Histogram of the imaginary part of $\zeta(\sigma)$
with different values of the damping obtained with method
(ii); (b) Histogram of the real
part of $\zeta(\sigma)$ with different dampings obtained with method
(ii).
(c) and (d) are histograms of the
reactance and resistance from
HFSS calculation with a lossy top and bottom plate, compared with
histograms from Eq.~(\ref{eq:xsig}) computed as in (a) and (b) (dashed
line) and by method (i) (solid line).
}
\label{fig:xsig} \end{figure}

In analogy to Eq.~(\ref{eq:corresp2}) we write for the cavity
impedance
\begin{equation}
Z=jX_R+R_R\zeta,  \label{eq:corresp3}
\end{equation}
and we use (\ref{eq:xsig}) to generate probability distribution
functions for the real and imaginary part of $\zeta=\rho+j\xi$.
We first generate $N$ values of $w_n$ as independent Gaussian
random variables of unit width (for this purpose we use a suitable
random number generator). We next generate $N$ values of the
normalized eigenvalues $\tilde k_n^2$. To do this we have utilized
two methods: (i) an approximate method based on
Eq.~(\ref{eq:spacetrs}) (for the TRS case) or
Eq.~(\ref{eq:spacetrsb}) (for the TRSB case), and (ii) a method
based on random matrix theory. We pick the value of $k^2$ relative
to the spectrum $k_n^2$ such that the median of $\xi$ is zero.

For method (i) we independently generate $N$ values of $s_n$ using
the distribution (\ref{eq:spacetrs}) or (\ref{eq:spacetrsb}). We
then obtain $\tilde{k}_n^2$ as $\tilde{k}_n^2=\sum_{n'=1}^{n}
s_{n'}$. The main assumption of this method is that the spacings
$s_n$ can be usefully approximated as uncorrelated. On the other
hand, it  is known from random matrix theory that the spacings are
correlated over long distance (in $n$), and the thus the
assumption of method (i) is questionable (compare (46) and (47)).
This motivates our implementation of method (ii) (See also
\cite{ekogan}).

To implement method (ii) we generate an $M\times M$ random matrix with $M$ large
($M$=1000) drawn from the appropriate ensemble (GOE or GUE) again using a
random
number generator. The width of the diagonal elements is taken to be unity. We then
numerically determine  the eigenvalues. The average spacing between eigenvalues of
random matrices is not uniform. Rather, in the limit of large $M$, the eigenvalues
$\lambda$ are distributed in the range $-\sqrt{2M}<\lambda<\sqrt{2M}$, and the
average spacing for eigenvalues near an eigenvalue $\lambda$ is given by
\begin{equation}
\Delta(\lambda)=\pi/\sqrt{2M-\lambda^2}
\label{eq:rm_leveldensity}
\end{equation}
in both the TRS and TRSB cases. In order to generate a sequence of eigenvalues
with approximately uniform spacing we select out the middle 200 levels. We then
normalize the eigenvalues by multiplying $1/\Delta(0)$ to create a
sequence
of $\tilde k_n^2$ values with average spacing of unity.

Histogram approximations to the GOE probability distributions of $Re[\zeta]$ and
$Im[\zeta]$ obtained by use of (\ref{eq:xsig}) and method (ii) are shown in
Figs. 7(a) and 7(b). These were obtained using 30,000 random GOE matrix
realizations of 1000 by 1000 matrices and selecting the middle 200 eigenvalues of
each realization. The resulting graphs are shown for a range of damping values,
$\tilde k^2\sigma$=0.01, 0.1, 0.5, 1, 5 and 10.  As
seen in Fig.~\ref{fig:xsig}(a), when $\tilde k^2\sigma$
is increased, the distribution of $\xi$ values becomes ``squeezed".
Namely, the Lorentzian tail disappears and the fluctuations in
$\xi$ decrease. Eventually, when $\sigma$ enters the regime, $1
\ll \tilde k^2\sigma \ll N$, the probability distribution function
 of $\xi(\sigma)$ approaches a narrow Gaussian distribution centered
 at $\xi =0$ (recall that $\bar \xi =0$).  As shown in
Fig.~\ref{fig:xsig}(b), as
$\sigma$ increases from zero, the distribution of the real part of
$\zeta(\sigma)$ which, for $\sigma=0$, is a delta function at
zero, expands and shifts toward 1,  becoming peaked around 1.
In the very high damping case, both the real part and imagnary parts of
$\zeta$, $\rho$ and $\xi$, will be Gaussian distributed with the mean
value equal to 1 and 0 respectively, and the same variance inversely
proportional to the loss (as shown in the Appendix). As a consequence,
the reflection coefficient $|S|^2$ in the high damping limit, is
exponentially distributed. This result is consistent with the
theoretical discussion given by \cite{ekogan}.

In general, the complex impedance distribution is not described
using  simple distributions such as Gaussian or Lorenzian. The
distribution of the real part of the impedance has been studied in
connection with the theory of mesoscopic systems and known as the
``local density of states"(LDOS). Through the supersymmetry
approach, Efetov obtained the probability density function  for
the LDOS in systems without time reversal symmetry\cite{efetov}.
For chaotic systems with  time reversal symmetry, the
corresponding exact formula was derived in a form of multiple
integral \cite{taniguchi}. However the difficulty to carry out the
five-fold integral makes it hard to interpret the formulus in
Ref.~\cite{taniguchi}. Very recently, Fyodorov and
Savin have proposed an interpolation formulus for the impedance
distributions at arbitary values of damping parameter
\cite{fyodorov04}. The suggested formulas satisfy all the asymptotic
behaviors  in the physically interesting limiting cases, e.g. weak damping
or very string damping cases. Furthermore, these formulas seem to agree
pretty well with the results of the numerical simulations, though the
agreement in the intermediate damping case is not as good as in the
limiting cases. In our paper, we  still use the histograms
generated from the Monte-Carlo
simulations as a comparison to the HFSS data, however, we believe the
formula presented by Fyodorov and Savin would be very helpful for the most
of practical purposes of comparison.

We noted that the variance of the real and imaginary parts of the
complex impedance are equal. There is a more
fundamental connection between these that is revealed by considering the
reflection coefficient in the perfectly
matched case,
\begin{equation}
\alpha e^{j\phi}=(\zeta-1)/(\zeta+1),
\label{eq:zeta2_inv}
\end{equation}
where $\alpha$ and $\phi$ are random variables giving the magnitude and
phase of the reflection coefficient. It can be argued \cite{ekogan} that
$\phi$ and $\alpha$ are independent and that $\phi$ is uniformly
distributed in [0, $2\pi$]. The magnitude $\alpha$ is distributed on the
interval [0, 1] with a density that depends on losses. A plot of the
probability distribution for $\alpha$ taken from the data in Figs
(7a) and (7b) is shown in Fig~8, for the damping values 0.1, 0.5, 1 and 5.

\begin{figure}
\includegraphics[scale=0.4]{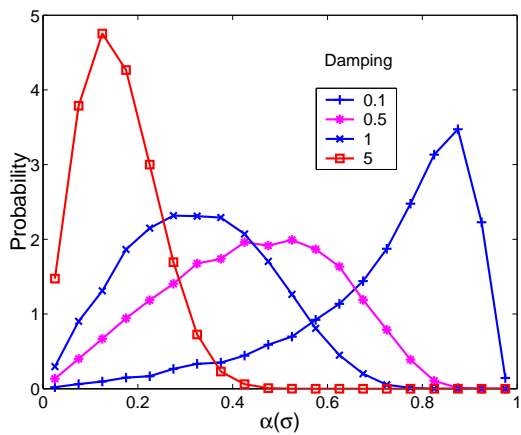}
\caption{Histogram of the magnitude of reflection coefficient in the
Eq.~(\ref{eq:zeta2_inv}), $\alpha(\sigma)$, with different values of the
damping.}
\end{figure}

We can express the actual complex reflection coefficient $\rho$ in terms
of the normalized reflection coefficient by first finding the normalized
impedance from (\ref{eq:zeta2_inv}), $\zeta=(1+\alpha e^{j\phi})/(1-\alpha
e^{j\phi})$ calculating the cavity impedance from (\ref{eq:corresp3}), and
expressing the result in terms of the radiation reflection coefficient
(\ref{eq:rhor}). The result is 
\begin{equation}
\rho=\frac{\rho_R+\alpha e^{j(\phi+\Delta\phi)}}{1+\alpha
e^{j(\phi+\Delta \phi)}\rho_R^{*}},
\label{eq:rhopc2rho}
\end{equation}
where $\tan(\Delta\phi/2)=-X_R/(R_R+Z_0)$ depends on system specific
parameters. Since the angle $\phi$ is uniformly distributed, it can be
shifted by $\Delta\phi$ thus eliminating $\Delta\phi$ from the
expression. Eq.~(\ref{eq:rhopc2rho}) is then a restatement of the Poisson
kernel in the single port case.

The independence of $\alpha$ and $\phi$ in Eq.~(\ref{eq:zeta2_inv}) also
guarantees the invariance of
the distribution of cavity impedances when a lossless two port is added
as in Sec III(C). In particular, the normalized cavity impedance $\zeta$
before the addition of the two port is given by
\begin{equation}
\zeta=\frac{Z-jX_R}{R_R}=\frac{1+\alpha e^{j\phi}}{1-\alpha e^{j\phi}}.
\label{eq:zeta2}
\end{equation}
With the addition of the lossless two port as shown in the Fig.~5,
impedances are transformed to $Z'$, $X_R'$, and $R_R'$ such that
\begin{equation}
\zeta=\frac{Z'-jX_R'}{R_R'}=\frac{1+\alpha e^{j(\phi-\phi_c)}}
{1-\alpha e^{j(\phi-\phi_c)}}.
\label{eq:zeta3}
\end{equation}
where $\phi_c=(2\beta+\pi)$ depends only on the properties of the two port and the
cavity coupling port and the angle $\beta$ satisfies
\begin{equation}
\begin{aligned}
\cos\beta &=\frac{R_R}{\sqrt{R_R^2+(X_{11}+X_R)^2}}, \\
\sin\beta &=\frac{(X_{11}+X_R)}{\sqrt{R_R^2+(X_{11}+X_R)^2}}.
\end{aligned}
\end{equation}
Since $\phi$ is
uniformly distributed,
so is the difference $\phi-\phi_c$. Consequently, the normalized random variables
$\zeta$ and $\zeta'$ have identical statistical properties.

A by-product of (\ref{eq:zeta2}) is that we can easily prove that its
real
part $\rho=(1-\alpha^2)/(1+\alpha^2-2\alpha \cos\phi)$ and its imaginary
part
$\xi=(2\alpha\sin\phi)/(1+\alpha^2-2\alpha\cos\phi)$ have the same
variance and zero correlation. Since $\alpha$ and $\phi$ are independent,
we can carry out
the
integration over the uniformly distributed $\phi$ and obtain
\begin{equation}
Var[\rho]=Var[\xi]=\langle
\frac{1+\alpha^2}{1-\alpha^2}\rangle_{\alpha}-1, \qquad Cov[\rho, \xi]=0
\end{equation}
where $\langle .. \rangle _{\alpha}$ denotes average over $\alpha$. This
property has been tested in microwave cavity experiments with excellent
agreements \cite{anlageprl}. For the high damping case, $\rho-1$ and
$\xi$
will become two independent Gaussian variables with zero mean and
small but same
variances. This yields an exponential distribution for the $\alpha ^2$,
which is consistent with the result obtained by Kogan \cite{ekogan} based
on the ``maximum information entropy" principle. For the weakly
absorpting case, Beenakker and Brouwer \cite{beenakker01} studied the
distribution of $\alpha^2$ in the TRSB case through
the time-delay matrix and obtained a generalized Laguerre ensemble.
 Howerver, for a TRS system
with arbitary loss,
there is no simple formula for the distribution of reflection
coefficients.

Using HFSS, we simulate the lossy case by specifying the  material
on the top and
bottom plates to be an
imperfect conductor with a bulk resitivity of
70 $m\Omega\cdot cm$. In this case we can calculate a value of
$\sigma=\delta/h=0.002$, where $\delta$ is the skin depth and $h$  the
cavity height. The corresponding parameter $\tilde{k}^2 \sigma$ is 0.5 at
7.75GHz. Histogram results for the normalized reactance ($\xi$)  and
resistance ($\rho$) fluctuations of
$\zeta_{hfss}=R_R^{-1}(Z_{cav}-jX_R)=\rho+j\xi$ are plotted in
Figs.~\ref{fig:xsig}(c) and \ref{fig:xsig}(d) together with the histograms
generated from
Eq.~(\ref{eq:xsig}),
and using spectra from the random matrices. As can be
seen, the histograms
from the HFSS simulations match those of the model.

\section{Summary}

We have applied the concepts of wave chaos to the problem of
characterizing the statistics of  the impedance and scattering coefficient for
irregular
electromagnetic cavities with one port in the small wavelength regime.  The
coupling of energy in and out of the port in such cavities depends on
both the geometry of the  port and the geometry of the cavity.  We
 found that  these effects can
approximately be separated.  The geometry of the port is
characterized by its radiation impedance which has both a real and
imaginary part.  This impedance describes the port in the case in
which the distant walls of the cavity are treated  as perfect
absorbers (or else are removed to infinity).  The effects of geometry of
the cavity can be treated
in a statistical way using Random Marix Theory. The separation of
the system specific aspects of the coupling and the universal
aspects has previously been described using the Poisson kernel
\cite{mello85}. The relation of our approach to the Poisson kernel 
may be understood by comparing the equivalent relations
(\ref{eq:corresp3}) and (\ref{eq:rhopc2rho}). To extract a universal
quantity ($\zeta$) from a set of impedance values ($Z$) one must subtract
the radiation reactance and normalize to the radiation resistance. To
extract a universal quantity ($\alpha e^{j\phi}$) from a  set of
reflection values ($\rho$) we must solve the bilinear relation
(\ref{eq:rhopc2rho}) for the magnitude $\alpha$ and phase $\phi$ of the
normalized reflection coefficient. The normalized impedance and scattering
amplitude are related by $\alpha e^{j\phi}=(\zeta-1)/(\zeta+1)$. If the
radiation reflection coefficient $\rho_R$ is known then
(\ref{eq:rhopc2rho}) may be solved directly for normalized reflection
coefficient,
\begin{equation}
\alpha e^{j\phi}=e^{-j\Delta\phi}\frac{\rho-\rho_R}{1-\rho \rho_R^{*}}.
\label{eq:rho2rhopc}
\end{equation}

The radiation reflection coefficient can be determined directly by
measurement \cite{anlageprl} or by ensemble averaging. According to
Ref.~\cite{mello85} the average of $\rho$ is equal to $\rho_R$. This can
be verified directly from (\ref{eq:rhopc2rho}) by averaging over the
uniformly distributed phase $\phi$. Regardless of the value of $\alpha$,
one finds 
\begin{equation}
\int \frac{d \phi}{2\pi}\rho= \rho_R.
\label{eq:rhoave}
\end{equation} 
Thus if enough appropriate, statistically independent realizations are
available to computer the average of $\rho$, (\ref{eq:rhoave}) can be used
to find the universal reflection amplitude.

 Consistent with previous results \cite{mello85} our model predicts
that in
the lossless case
the impedance is Lorentzian distributed
with a mean equal to the radiation reactance and a width equal to
the radiation resistance.
The Lorentzian prediction is tested by direct numerical solution
of Maxwell's equation for the cavity of Fig.~\ref{fig:bowtie}.
The predictions are verified if an additional averaging over
frequency is introduced. Effects of  distributed loss and variation of
coupling are also investigated and we have generated pdf's for the real
and
imaginary parts of the normalized impedance. In addition, we have
calculated the mean and variance for these distributions and determined
the effect of correllations in the eigenfrequencies on the variances. The
values of the variance depend on the degree of loss in the cavity and can
be used to qunatify it. Finally, we have compared the predicted
distributions of the normalized impedance with those obtained
from a direct numerical simulation.

\begin{acknowledgments}
We thank R. E. Prange, S. Fishman, J. Rogers and S. Anlage for
discussions and help. This work was supported in part by the DOD
MURI for the study of microwave effects under AFOSR Grant
F496200110374.
\end{acknowledgments}
\newpage
\appendix*
\section{Variance of Cavity Reactance and Resistance in the Lossy Case. }

From Eq.~(\ref{eq:zfinal_loss}), we obtain the expression for the
complex
impedance in the single port case,
\begin{equation}
\begin{aligned}
Z(\sigma)
&=\frac{1}{\pi}\sum_1^N [\frac{\Delta(k_n^2) R_R(k_n^2) w_n^2
[k_d^2+j(k_n^2-k^2)]}{(k^2-k_n^2)^2+(k_d^2)^2}] \\
&=R(\sigma)+jX(\sigma),
\end{aligned}
\label{eq:C1}
\end{equation}
where $\Delta$ is the mean spacing $\langle
k_n^2-k_{n-1}^2\rangle$, $X(\sigma)$ and $R(\sigma)$ are cavity reactance
and resistance in the lossy case and $k_d^2=k^2\sigma$. In this appendix,
we are going to
evaluate the mean and variance of $X(\sigma)$ and $R(\sigma)$ as well as
their covariance.

We first investigate the mean of $R(\sigma)$. We express the mean in terms
of
probability distribution function for the eigenvalues $k_n^2$.
\begin{equation}
\begin{aligned}
E[R(\sigma)]&=\frac{1}{\pi}\int\dots\int
dk_1^2 \dots dk_N^2
P_J(k_1^2,\ldots , k_N^2) \\
& \sum_{n'=1}^{N}\frac{R_R \Delta \langle
w_{n'}^2
\rangle
k_d^2}{(k^2-k_{n'}^2)^2+k_d^4},
\end{aligned}
\label{eq:eRsig}
\end{equation}
where $P_J$ is the joint distribution of eigenlevels
($k_1^2,\ldots, k_N^2$) assuming they are unordered. Since the levels are
not ordered, in each term of the sum,
we can integrate over all $k_n^2\neq k_{n'}^2$,
and obtain $N$ identical terms. Thus,
\begin{equation}
E[R(\sigma)]=\frac{N}{\pi}\int dk_{n'}^2 P_1(k_{n'}^2)R_R\Delta \langle w^2
\rangle \frac{k_d^2}{(k^2-k_{n'}^2)^2+k_d^4}
\label{eq:eRsig_2}
\end{equation}
where $P_1(k_{n'}^2)$ is distrubtion for a single level. Here we have
introduced an integer $N$ representing the total number of levels. We
will take the limit of $N \rightarrow \infty$. The single level
probability distribution then satisfies by definition,
\begin{equation}
P_1(k_{n'}^2)=\frac{1}{N\Delta(k_{n'}^2)}.
\label{eq:P1}
\end{equation}
We next assume that the radiation resistance $R_R(k_{n'}^2)$ is relatively
constant over the interval of $k_{n'}^2$ values satisfying
$|k^2-k_{n'}^2|<k_d^2$ and we will move it outside the integral
replacing it by $R_R(k^2)$. Assuming that $k_d^2$ is not too large
($k_d^2\ll
k^2$)  we
can take the end points at the integral to plus and minus infinity and
evaluate Eq.~(\ref{eq:eRsig_2}) as
\begin{equation}
E[R]=\frac{R_R}{\pi} \int_{-\infty}^{\infty} dx \frac{1}{x^2+1}=R_R(k^2),
\end{equation}
where $x=(k_{n'}^2-k^2)/k_d^2$. Thus the expected value of the real part
of cavity impedance is the radiation resistance independent of the amount
of damping. This is somewhat surprising since we have previously asserted
that in the lossless case, the cavity resistance is zero. The constancy of
the expected resistance results from the resonant nature of the cavity
impedance. When losses are small, $k^2\sigma=k_d^2 \ll 1$, for almost all
frequencies the resistance is small.  However, for the small set of the
frequencies near a resonance the resistance is large. This is evident in
the histograms of Fig.~(\ref{fig:xsig}b). The result is that small chance
of a large
resistance and large chance of small resistance combine to give an expected
value
resistance which is constant.

In order to obtain the variance of $R(\sigma)$, we  calculate the
second moment of
$R(\sigma)$,
\begin{equation}
\begin{aligned}
E[R(\sigma)^2] &=(\frac{1}{\pi})^2
\int\dots \int\ dk_1^2 \dots dk_N^2
P_J(k_1^2,\ldots,
k_N^2) \\
& \sum_{n',m'=1}^{N}\frac{\Delta^2 R_R(k_{n'}^2) R_R(k_{m'}^2)
\langle w_{m'}^2 w_{n'}^2 \rangle  k_d^4}
{((k^2-k_{m'}^2)^2+k_d^4) ((k^2-k_{n'}^2)^2+k_d^4)}\\
&\equiv
I_1+I_2.
\end{aligned}
\label{eq:eR2sig_1}
\end{equation}
Following the arguments advanced to calculate $E[R(\sigma)]$, we note that
there will be $N$ terms in the double sum for which $k_{n'}^2=k_{m'}^2$
giving
\begin{equation}
I_1=\frac{N}{\pi^2}\int dk_{n'}^2 P_1(k_{n'}^2)
\frac{\Delta^2 R^2(k_{n'}^2)\langle w_{n'}^4 \rangle k_d^4}
{[(k^2-k_{n'}^2)^2+k_d^4]^2}
\label{eq:eR2_I1}
\end{equation}
and $N(N-1)$ terms for which $k_{m'}^2\neq k_{n'}^2$ giving
\begin{equation}
\begin{aligned}
I_2&=N(N-1)\iint dk_{n'}^2 dk_{m'}^2\\
&\frac{P_2(k_{n'}^2,k_{m'}^2)\Delta (k_{n'}^2) \Delta(k_{m'}^2
R_R(k_{n'}^2) R_R(k_{m'}^2) \langle w_{n'}^2 \rangle
\langle w_{m'}^2 \rangle k_d^4}
{[(k^2-k_{n'}^2)^2+k_d^4] [(k^2-k_{m'}^2)^2+k_d^4]}.
\end{aligned}
\label{eq:eR2_I2}
\end{equation}
For the first integral we use (\ref{eq:P1}) for the single level
distribution function, and making the same approximation as before,
we
obtain
\begin{equation}
I_1=R_R^2(k^2)\frac{\langle w^4 \rangle \Delta(k^2)}{2\pi k_d^2}.
\label{eq:eR2_I1_result}
\end{equation}
For the second integral we need to introduce the two level
distribution
function. For the spectra that we consider, the two level
distribution has
the form
\begin{equation}
P_2(k_{n'}^2,k_{m'}^2)=(\frac{1}{N\Delta})^2[1-g(|k_{n'}^2-k_{m'}^2|)].
\label{eq:P2}
\end{equation}
Here the function $g$ describes the correllations between two energy
levels. For uncorrellated levels giving a Poisson distribution of
spacings
we have $g=0$. In the presence of level repulsion we expect $g(0)=1$
with
$(1-g)\propto |k_{n'}^2-k_{m'}^2|^{\beta}$ for small spacing, and
$\beta=1$ for TRS and $\beta=2$ for TRSB systems. As
$|k_{n'}^2-k_{m'}^2|\rightarrow \infty$, $g\rightarrow 0$ indicating
loss
of correllation for two widely separated levels. The function $g$
will be
different for spectra produced by random matrices and spectra
genergated
from sequences of independent spacings. Expressions of $g$ for the
spectra
of random matrices can be found in the book by Mehta
(\cite{mehta91},
Ch. 5 \& 6). We will derive the expression for $g$ for
spectra generated by sequences of independent spacings later in this
appendix.

Based on expression (\ref{eq:P2}) and the  usual assumptions on the
slow variations of $R_R$ and $\Delta$ with eigenvalue $k_{n'}^2$ we obtain
\begin{equation}
I_2=(E[R])^2-I_g,
\label{eq:eR2_I2_Ig}
\end{equation}
where the first term comes from the $1$ in \ref{eq:P2} and the second term
comes from the correllation function g
\begin{equation}
I_g=\frac{R_R(k^2)\langle w^2 \rangle ^2}{\pi} \int_{-\infty}^{\infty}
\frac{d \tilde{k}^2}{k_d^2}
\frac{2}{4+(\tilde{k}^2/k_d^2)^2}g(|\tilde{k}^2|).
\label{eq:eR2_Ig}
\end{equation}
The variance of $R$ is thus given by
\begin{equation}
\begin{aligned}
Var[R]&=E[R]^2-E[R^2]\\
&=\frac{R_R^2}{\pi}\frac{\Delta}{k_d^2}[\frac{\langle w^4\rangle}{2}-
\langle w^2 \rangle ^2 \int_{-\infty}^{\infty}
\frac{d\tilde{k}^2}{\Delta}\frac{2
g(|\tilde{k^2}|)}{4+(\tilde{k}^2/k_d^2)^2}].
\label{eq:VarR_1}
\end{aligned}
\end{equation}
Note, since $w$ is a Gaussian random variable with zero mean and unit
variance, $\langle w^2 \rangle =1$ and $\langle w^4 \rangle =3$.

Equation (\ref{eq:VarR_1}) shows that the variances of $R$ depends on
$k_d^2/\Delta$, the ratio of the damping width to the mean spacing of
eigenvalues. In the low damping case, $k_d^2/\Delta \ll 1$, the integrand
in (\ref{eq:VarR_1}) is dominated by the values of $|\tilde{k}^2|<\Delta$
and we replace $g$ by its value $g(0)$. Doing the integral we find
\begin{equation}
Var[R]=R_R^2[\frac{\Delta}{k_d^2}\frac{\langle w^4 \rangle}{2\pi}-
g(0)\langle w^2 \rangle ^2].
\label{eq:VarR_2}
\end{equation}
Since the damping is small, the first term dominates and the variance is
independent of the eigenvalue correllation function. This is consistent
with our previous findings that the eigenvalue statistics did not affect
the
distribution of reactance values.

In the high damping limit, $k_d^2>\Delta$, the
integral in (\ref{eq:VarR_1}) is dominated by $\tilde k^2$ values of order
$\Delta$ and we have,
\begin{equation}
Var[R]=\frac{R_R^2}{\pi}\frac{\Delta}{k_d^2}[\frac{3}{2}-\int_{0}^{\infty}\frac{d\tilde
k^2}{\Delta} g(|\tilde k^2|)].
\label{eq:VarR_3}
\end{equation}
The variance decreases as damping increases with a coefficient that
depends on the correllation function. Physically the correllations are
important because in the high damping case a relatively large number of
terms in the sum (\ref{eq:C1}) contribute to the impedance and the sum is
sensitive to correllations in these terms.

The integral of the correllation function can be evaluated for different
spectra. For spectra generated from random matrices, we have
(\cite{mehta91}, Ch.6)
\begin{equation}
g(s)=f(s)^2-\frac{\partial f}{\partial s}
[(\int_0^s  d s' f(s'))-\frac{1}{2}sgn(s)]
\label{eq:corr_goe}
\end{equation}
for TRS matrices and
\begin{equation}
g(s)=f(s)^2
\label{eq:corr_gue}
\end{equation}
for TRSB matrices, where $f(s)=\sin(\pi s)/(\pi s)$. In both cases,
we find
\begin{equation}
\int_0^{\infty}ds g(s)=\frac{1}{2}.
\label{eq:int_ge}
\end{equation}
However, to consider the TRSB case we need to repeat the calculation
including complex values of the Gaussian variable $w$. The result is
\begin{equation}
Var[R(\sigma)]=\frac{R_R^2}{\pi}\frac{\Delta}{k_d^2}
[1-\int_0^{\infty}\frac{d\tilde{k}^2}{\Delta}g(|\tilde{k}^2|)].
\label{eq:VarR_gue}
\end{equation}
For spectra generated by sequences of independent spacing distributions
we will show
\begin{equation}
\int_0^{\infty}\frac{d\tilde{k}^2}{\Delta}g(|\tilde{k}^2|)=1-\frac{1}{2}
\langle s^2 \rangle,
\label{eq:Int_g_indep}
\end{equation}
where $\langle s^2 \rangle$ is the expected value for the normalized
nearest
neighbor spacing squared. Using (\ref{eq:spacetrs}) and
(\ref{eq:spacetrsb}), this
gives
\begin{equation}
\int_0^{\infty}\frac{d\tilde{k}^2}{\Delta}g(|\tilde{k}^2|)=
  \begin{cases}
    1-\frac{2}{\pi} & \qquad \text{for TRS}, \\
    1-\frac{3\pi}{16} & \qquad \text{for TRSB}.
  \end{cases}
\label{eq:eq:Int_g_indep_2}
\end{equation}
Note also that (\ref{eq:Int_g_indep}) gives the required value of zero
for Poisson spacing distributions, where $\langle s^2 \rangle =2$.

We can evaluate the
expected value of the reactance and its variance, as well as the covariance
of
reactance and resistance, using the same approach. We find the expected
value of
reactance is given by the radiation reactance,
\begin{equation}
E[X]=X_R(k^2).
\label{eq:mean_Xsig}
\end{equation}
The variance of the reactance is equal to that of the resistance
(\ref{eq:VarR_1})
the covariance between them is zero.

We now derive the $g$-integral (\ref{eq:Int_g_indep}) for spectra
generated from
independent spacings. We introduce a  conditional distribution $P_m(s)$
that is
 the probability density that the $m^{th}$ eigenvalue is in the range
[$s$, $s+ds$] given that eigenvalue $m=0$, is at zero.
For convenience, here $s$ is the normalized spacing with unit mean. When
$m=1$,
$P_1(s)$ is the spacing distribution $p(s)$.
    Thus, $1-g(s)$ stands
for the probability that there exists an eigenlevel at [s, s+ds] given
one level located at 0. This equality  can be expressed as the
summation of $P_m(s)$,
\begin{equation}
1-g(s)=\sum_{m=1}^{\infty}P_m(s).
\end{equation}
$P_m(s)$ can be evaluated assuming the spacings are independent,
\begin{equation}
1-g(s)=\sum_{m=1}^{\infty}[\int
\prod_{i=1}^{m} ds_i P_1(s_i)\delta(s-\sum_{i=1}^m s_i)].
\label{eq:gs}
\end{equation}
We  Laplace transforme both sides of Eq.~(\ref{eq:gs}), and obtain
\begin{equation}
\frac{1}{\tau}-\int_0^{\infty}ds e^{-\tau
s}g(s)=\sum_{m=1}^{\infty}[\bar{P}_1(\tau)]^m=\frac{\bar{P}_1(\tau)}
{1-\bar{P}_1(\tau)}.
\end{equation}
To evaluate $\int_0^{\infty} ds g(s)$, we take the limit of $\tau
\rightarrow 0$.
The transform $\bar{P}_1(\tau)$ can be expressed in terms of the moments of
$P_1(s)$,
\begin{equation}
\begin{aligned}
\bar{P}_1(\tau)&=\int_{0}^{\infty}e^{-s\tau}P_1(s)ds,\\
& \sim \int_{0}^{\infty}(1-s\tau+\frac{s^2 \tau ^2}{2})P_1(s)ds,\\
&=1-\tau \langle s\rangle + \frac{\tau^2}{2}\langle s^2 \rangle.
\end{aligned}
\end{equation}
Thus, we can evaluate the integration of $g(s)$ to be
\begin{equation}
\begin{aligned}
\int_{0}^{\infty}ds g(s)&=\lim_{\tau\rightarrow 0}\int_{0}^{\infty}ds
e^{-\tau s}g(s)\\
&=\lim_{\tau\rightarrow 0}
[\frac{1}{\tau}-\frac{\bar{P}_1(\tau)}{1-\bar{P}_1(\tau)}]\\
&=1-\frac{1}{2}\langle s^2 \rangle,
\end{aligned}
\end{equation}
which is Eq.~(\ref{eq:Int_g_indep}).

\end{document}